\documentclass[12pt,a4]{article}
\usepackage[utf8]{inputenc}
\usepackage[english]{babel}
\renewcommand{\baselinestretch}{1.25}

\usepackage{amsthm,amsmath,latexsym,amssymb,amsfonts,amscd}
\usepackage{graphics,lscape,fancyhdr,array,stmaryrd,euscript,wrapfig}
\usepackage{ifthen,empheq,slashed}
\usepackage{ifpdf}
\usepackage{verbatim,slashed}
\numberwithin{equation}{section}
\usepackage{hyperref,setspace}

\usepackage{mathrsfs}
\usepackage[numbers,sort&compress]{natbib}
\usepackage[nottoc]{tocbibind}

\newcommand{\pl}{\partial}

\newcommand{\be}{\begin{equation}}
\newcommand{\ee}{\end{equation}}

\newcommand{\adD}{{\mathsf{D}}}
\newcommand{\tadD}{\widetilde{\mathsf{D}}}

\newcommand{\Fron}{{\Phi}}

\newcommand{\mm}{{\ensuremath{\underline{m}}}}
\newcommand{\nn}{{\ensuremath{\underline{n}}}}

\newcommand{\aAs}{{\ensuremath{\mathsf{A}}}}

\newcommand{\aA}{{\ensuremath{\mathcal{A}}}}
\newcommand{\aB}{{\ensuremath{\mathcal{B}}}}

\newcommand{\ga}{\alpha}
\newcommand{\gb}{\beta}
\newcommand{\gc}{\gamma}

\newcommand{\gad}{{\dot{\alpha}}}
\newcommand{\gbd}{{\dot{\beta}}}
\newcommand{\gdd}{{\dot{\gamma}}}

\newcommand{\gnd}{{\dot{\nu}}}

\newcommand{\bry}{{{\bar{y}}}}
\newcommand{\brz}{{{\bar{z}}}}

\newcommand{\klein}{{\varkappa}}
\newcommand{\brklein}{{\bar{\varkappa}}}

\newcommand{\fud}[2]{{}^{#1}{}_{#2}\,}
\newcommand{\fdu}[2]{{}_{#1}{}^{#2}\,}

\DeclareMathOperator{\sign}{sign}

\newcommand{\pb}{{\bar{p}}}

\newcommand{\omegatwo}{\omega^{(2)}}
\newcommand{\Ctwo}{C^{(2)}}
\newcommand{\omegaone}{\omega^{(1)}}
\newcommand{\Cone}{C^{(1)}}

\newcommand{\Uf}{{\boldsymbol{U}}}
\newcommand{\Gf}{{\boldsymbol{G}}}
\newcommand{\Wf}{{\boldsymbol{W}}}
\newcommand{\Kf}{{\boldsymbol{K}}}
\newcommand{\Sf}{{\boldsymbol{S}}}
\newcommand{\Zf}{{\boldsymbol{Z}}}
\newcommand{\Df}{{\boldsymbol{\widetilde{D}}}}
\newcommand{\Dfa}{{\boldsymbol{D}}}
\newcommand{\Rf}{{\boldsymbol{R}}}
\newcommand{\Ff}{{\boldsymbol{R}}}
\newcommand{\Qf}{{\boldsymbol{Q}}}
\newcommand{\Pf}{{\boldsymbol{P}}}
\newcommand{\Bf}{{\boldsymbol{B}}}
\newcommand{\str}{{\mathrm{str}}}

\newcommand{\besubeqs}{\begin{subequations}}
\newcommand{\esubeqs}{\end{subequations}}

\begin{document}
\hfill
\begin{flushright}
    {LMU-ASC 12/17}
\end{flushright}
\vskip 0.015\textheight
\begin{center}

{\Large\bfseries  Formal Higher-Spin Theories \\
\vspace{0.2cm}
and\\
\vspace{0.4cm}
Kontsevich-Shoikhet-Tsygan Formality} \\

\vskip 0.03\textheight

Alexey \textsc{Sharapov}${}^{1}$ and Evgeny \textsc{Skvortsov}${}^{2,3}$

\vskip 0.03\textheight

{\em ${}^{1}$Physics Faculty, Tomsk State University, \\Lenin ave. 36, Tomsk 634050, Russia}\\
\vspace*{5pt}
{\em ${}^{2}$ Arnold Sommerfeld Center for Theoretical Physics\\
Ludwig-Maximilians University Munich\\
Theresienstr. 37, D-80333 Munich, Germany}\\
\vspace*{5pt}
{\em ${}^{3}$ Lebedev Institute of Physics, \\
Leninsky ave. 53, 119991 Moscow, Russia}

\vskip 0.02\textheight

{\bf Abstract }

\end{center}

\begin{quotation}
The formal algebraic structures that govern higher-spin theories within the unfolded approach turn out to be related to an extension of the Kontsevich Formality, namely, the  Shoikhet--Tsygan Formality. Effectively, this allows one to construct the Hochschild cocycles of higher-spin algebras that make the interaction vertices. As an application of these results we construct a family of Vasiliev-like equations that generate the Hochschild cocycles with $sp(2n)$ symmetry from the corresponding cycles. A particular case of $sp(4)$ may be relevant for the on-shell action of the $4d$ theory. We also give the exact equations that describe propagation of higher-spin fields on a background of their own. The consistency of formal higher-spin theories turns out to have a purely geometric interpretation: there exists a certain symplectic invariant associated to cutting a polytope into simplices, namely, the Alexander-Spanier cocycle.
\end{quotation}

\newpage

\renewcommand{\baselinestretch}{1.1}\normalsize
\tableofcontents
\renewcommand{\baselinestretch}{1.25}\normalsize

\section{Introduction}
\label{sec:}

Despite the conceptual simplicity of higher-spin theories --- in most of the cases they can be thought of as AdS/CFT \cite{Maldacena:1997re,Gubser:1998bc,Witten:1998qj} duals of free conformal fields theories \cite{Klebanov:2002ja,Sundborg:2000wp,Sezgin:2002rt} with the possibility of having more interesting dualities resulting from alternate or mixed boundary conditions \cite{Sezgin:2003pt,Leigh:2003gk,Giombi:2011kc,Chang:2012kt} --- many bulk questions are still to be answered. For example, there is no yet any proof of quantum consistency of higher-spin theories, though they are expected to be quantum finite, see e.g. \cite{Giombi:2014yra,Giombi:2014iua,Giombi:2013fka} for some one-loop results. There are many higher-spin theories that are expected to exist but have not yet been constructed in any form. One more obstacle is the lack of a standard\footnote{A non-standard action was proposed in \cite{Boulanger:2011dd}, a feature being that it does not have the Fronsdal kinetic terms. There are attempts to reconstruct higher-spin theories directly from the CFT side, e.g. \cite{Leigh:2014qca}, \cite{Koch:2014aqa}. This way the cubic action was found in \cite{Sleight:2016dba}, see also \cite{Bekaert:2014cea,Skvortsov:2015pea} and \cite{Kessel:2015kna} for the special case of $AdS_3$. A part of the on-shell quartic action in $d=4$ was reconstructed in \cite{Bekaert:2015tva}.} action principle for any of the higher-spin theories that one would attempt to quantize. Nevertheless, higher-spin theories should be the simplest toy models teaching us that higher-spin fields, whether massless or massive like in string theory, are important for quantization of gravity. Study of any of the problems listed above should be important for achieving this goal.

Historically, higher-spin theories were attacked directly from the bulk side \cite{Fradkin:1986qy,Fradkin:1986ka,Vasiliev:1988sa} long before the AdS/CFT era had begun. Since the earlier AdS/CFT papers addressing higher-spin theories it has been clear that knowing a CFT dual, i.e. just a free CFT, should be sufficient for constructing its bulk higher-spin dual, see e.g. recent \cite{Bekaert:2014cea,Bekaert:2015tva,Sleight:2016dba}. The precise relation between the CFT side and the structures underlying higher-spin theories remains unclear and this is one of the questions we address here. Specifically, we investigate the formal algebraic structures behind higher-spin theories within the unfolded approach \cite{Vasiliev:1988sa} or, what is the same, within the formal Q-manifold geometry. 

The main advantage of the unfolded approach is that it allows one to treat higher-spin symmetries in the exact way without having to perform weak-field expansion over any specific gravitational background like $AdS$. The expansion parameter is the deviation of a higher-spin connection from being flat. Therefore, every interaction term added to unfolded equations contributes to all orders in terms of the weak-field expansion. 

Unfolding is not a panacea  and some of the questions may still be difficult to answer. Knowing equations of motion certainly implies some information about the action, but cannot fully replace it, especially when the issue of quantization is addressed \cite{Kazinski:2005eb,Lyakhovich:2005mk,Kaparulin:2011zz,Kaparulin:2011aa}. It is still an open problem of how to reproduce at least all tree-level AdS/CFT correlators from the bulk, for which the equations of motion suffice \cite{Giombi:2009wh,Giombi:2010vg}. Another feature that has come to light recently \cite{Prokushkin:1999xq,Prokushkin:1998bq,Giombi:2009wh,Boulanger:2015ova,Skvortsov:2015lja,Taronna:2016ats} is that formally consistent unfolded equations can be ill-defined as differential equations. In particular, they can easily encode vertices/redefinitions that are forbidden by locality in field theory. 

Nevertheless, the unfolded formulation does uncover the structures that stay nontrivial even under a cavalier treatment of locality. It would be appropriate to refer to such structures as formal, which is justified by the need to have more ingredients as to control locality, construct and quantize actions. Also, as we explain, such structures can systematically be  generated via formality theorems. Formal higher-spin theories is what we would like to study in the paper and the qualifier `formal' may precede most of the statements made below.  

The sketch of the proposal is as follows. Any free CFT is fully determined by specifying a representation of the conformal algebra that the fundamental field(s) belong to. The higher-spin algebra associated to such a CFT is just the algebra of matrix elements in this representation or the algebra of symmetries of the corresponding conformally-invariant equation, e.g. $\square \phi=0$, \cite{Eastwood:2002su}. The same algebra can also be understood as a quantization of the coadjoint orbit corresponding to the fundamental field, see e.g. \cite{Michel2014,Joung:2014qya}. Higher-spin algebras come as associative algebras by construction. The simplest examples of higher-spin algebras are just the Weyl algebras with generators satisfying $[p_i,q^j]=\delta^i_j$, \cite{Vasiliev:1986qx}. Therefore, higher-spin algebras can be constructed within the deformation quantization approach \cite{Fedosov:1994zz,Kontsevich:1997vb}, with, for example, the Moyal-Weyl star-product being the trivial consequence of the Kontsevich formality theorem (which does not yet give a higher-spin theory). Higher-spin algebras are usually rigid and in any event it is not a deformation of the algebra structure that leads to formal higher-spin theories. It turns out that the first-order deformations are governed by certain Hochschild cocycle of higher-spin algebras. The cocycles can be constructed by employing an extension of the Kontsevich formality, namely, the Shoikhet-Tsygan formality theorem \cite{Tsygan,Shoikhet:2000gw}, which for the case of the Weyl algebra was explicitly done by Feigin, Felder and Shoikhet \cite{FFS}. 

We begin with a detour into the unfolded approach and its application to higher-spin theories, which was laid down by Vasiliev in \cite{Vasiliev:1988sa}. The basic model is the $4d$ formal higher-spin theory \cite{Vasiliev:1990vu,Vasiliev:1999ba}, but the statements are general enough. We show that the possibility to have global symmetries one the CFT side implies via Morita invariance that the interaction vertices are determined by the Hochschild cohomology of higher-spin algebras. Importantly, higher cohomology groups are expected to be trivial, which implies that the only non-trivial vertex is the first one and there are no obstructions at higher orders. Therefore, any formal higher-spin theory is fully determined by a single Hochschild cocycle of the relevant higher-spin algebra.

In the $4d$ case the first-order deformation is governed by a remarkable cocycle of the smallest Weyl algebra $A_1$, $[p,q]=1$, which was implicitly found in \cite{Vasiliev:1988sa}. Therefore, from the formal point of view the $4d$ Vasiliev equations provide a non-linear completion of the deformation induced by the $A_1$ Hochschild cocycle. There is a physically important, but mathematically inessential, detail that one needs to have two such independent deformations at a time --- one copy would lead to the holomorphic truncation of the theory, which was studied in \cite{Iazeolla:2007wt}. 

The Weyl algebra in $2n$ generators, which is denoted $A_{n}$, contains $sp(2n)$ subalgebra and is relevant for higher-spin theories and M/string-theory at least for few small $n$, see e.g. \cite{Bandos:1998vz,Vasiliev:2001dc,Plyushchay:2003gv,Bandos:2005mb} and \cite{Bergshoeff:2000qu,Gunaydin:1998bt,Gunaydin:1998km}. For example, the multiplet of all massless fields in $d=4$ enjoys $sp(8)$ symmetry, \cite{Fronsdal:1985pd,Vasiliev:2007yc}. As an application of the ideas above we construct a class of equations that take advantage of the $A_{n}$ Hochschild cocycles for any $n>1$, which is done in Section \ref{sec:HigherHS}. This allows us to shed light on the Vasiliev construction, where one of the crucial steps is to embed the deformation as a trivial one in a larger space. Similar construction works for all $A_{n}$. Therefore, we show how simple equations can generate highly nontrivial Hochschild cocycles. Mathematically, we construct a resolution of the Hochschild complex.

From the point of view of $A_n$ it becomes clear that the case of $A_1$ plays a special role and some of the fields, which are clearly different for $n>1$, can be identified for $n=1$. This identification yields the Vasiliev equations, which also bring in additional nonlinearities in the perturbation theory. Another feature is that the equations make sense without such an identification and provide description of fluctuations of higher-spin fields over a background of their own. It is noteworthy that one can directly jump from the equations for linear fluctuations over a sufficiently general background to the full nonlinear equations. In addition, the equations for the fluctuations allows one to write down the action of the global higher-spin transformations on the HS fields, with the Hochschild cocycle playing an important role.

It is interesting that the consistency of formal higher-spin theories can be understood geometrically as a possibility to cut a polytope made of $2n+2$ vertices in $2n$-dimensional space into $2n+2$ simplices. Dual to the Hochschild cocycle is an $sp(2n)$-invariant Alexander-Spanier cocycle that roughly speaking checks whether the origin belongs to a simplex, which for the case of $sp(2)$ was found in \cite{Vasiliev:1989xz}.  

We conclude in Section \ref{sec:Conclusions} with some comments and open problems. The main sections are supplemented with appendices which either provide more technical details or reformulate some of the sloppy statements in the main text in a mathematically rigorous way.

\section{Higher-Spins and Unfolding}
\label{sec:HSUnfolding}
By higher-spin theories we understand field theories that are non-linear completions of actions or at the very least equations of motion that describe free higher-spin fields when interactions are switched off.\footnote{It is a commonly accepted fact that the field content of such theories should have massless fields of unbounded spin, i.e. the multiplet is always infinite, which is easy to justify from the AdS/CFT point of view or by directly studying higher-spin symmetries \cite{Fradkin:1986ka,Maldacena:2011jn,Alba:2013yda,Boulanger:2013zza,Stanev:2013qra,Alba:2015upa}.} It is customary to represent free higher-spin fields by Fronsdal fields \cite{Fronsdal:1978vb}, with equations and gauge symmetries having the following schematic form:\footnote{$\mm,\nn,...=0,...,d-1$ are the base manifold indices, which, for free Fronsdal fields, is Minkowski or anti-de Sitter space. $\nabla_\mm$ is the associated covariant derivative. In most of the paper these indices will be hidden by contracting them with $dx^\mm$ and play no role.}
\begin{align}
\square  \Fron_{\mm_1...\mm_s}+...&=0\,, & \delta \Fron_{\mm_1...\mm_s}&= \nabla_{\mm_1} \epsilon_{\mm_2...\mm_s}+\text{permutations}\,.
\end{align}
It turned out to be a fruitful direction to replace Fronsdal fields by higher-spin vielbeins or frame fields \cite{Vasiliev:1980as}, which generalizes the metric vs. vielbein approaches to higher-spin fields,\footnote{$a,b,...=0,...,d-1$ are the fiber indices where the metric is constant metric $\eta^{ab}$ of the Lorentz algebra $so(d-1,1)$.}
\begin{align}
\Fron_{\mm_1...\mm_s}\Longrightarrow e^{a_1\ldots a_{s-1}}_\mm\, dx^\mm\,.
\end{align}
When supplemented with the higher-spin analogs of the spin-connection and summed over an appropriate range of spins, which is infinite $s=1,2,3,....$, the set of higher-spin frame fields turns out to form a single gauge connection $\omega$ of a higher-spin algebra \cite{Fradkin:1986ka}. Higher-spin gauge connection $\omega$ is a natural object from the symmetry point of view and via the gauge transformations $\delta \omega= d \xi-[\omega,\xi]$ it already knows more about interactions than the free Fronsdal fields do. In particular, one can construct actions that are consistent up to the cubic order just by using this symmetry, the Fradkin-Vasiliev actions \cite{Fradkin:1986qy}. However, at higher orders the symmetry needs to be deformed and this is a problem that can be addressed via an extension of the frame-like approach --- the unfolded approach \cite{Vasiliev:1988sa}, or, what is almost the same, the $Q$-manifold approach, and we review the two below. Next, we specialize to the simplest higher-spin theory of interest, the so-called Type-A model that contains fields with spins $s=0,1,2,3,...$.

\subsection{$Q$-manifolds and Formal Unfolding}
\label{sec:Qman}
The unfolded approach was introduced in \cite{Vasiliev:1988sa} as an extension of the frame-like approach and of the free differential algebras \cite{Sullivan77} approach to supergravities \cite{D'Auria:1982pm, Nieuwenhuizen:1982zf}. Mathematically, it is the same as $Q$-manifold \cite{Barnich:2005ru}, i.e. a graded manifold equipped with an odd  vector field $Q$ squaring to zero \cite{Kontsevich:1997vb}. Such vector fields are called usually {\it homological}.  In practice, the idea is to write equations in the form where the de Rham differential $d=dx^\mm \frac{\pl}{\pl x^\mm}$ of every field is expressed in terms of wedge-products of some other fields:
\begin{align}\label{dWQ}
&dW^\aA=Q^{\aA}(W)\,, && Q^{\aA}(W)=\sum_k Q^\aA_{\aB_1...\aB_k}W^{\aB_1}\wedge ...\wedge W^{\aB_k}\,.
\end{align}
Here $W^\aA$ is some, possibly infinite, set of fields that are differential forms of various degrees over the space-time manifold. Functional $Q^{\aA}(W)$ is assumed to have an expansion in terms of wedge products of the fields and $Q^\aA_{\aB_1...\aB_k}$ are space-time independent structure constants. The simplest example of an unfolded system would be a flat connection of some Lie algebra, where $Q^\aA_{\aB_1\aB_2}$ turns out to literally be the structure constants and $W^\aA$ are one-forms valued in the algebra. 

The formal consistency of the equations implies the Jacobi-like quadratic relations for the structure constants: \begin{align}\label{GrandJacobi}
0\equiv ddW^\aA&=dQ^{\aA}(W)=dW^\aB \wedge\frac{\overrightarrow{\pl} Q^\aA(W)}{\pl W^\aB} && \Longrightarrow&& Q^\aB\wedge\frac{\overrightarrow{\pl} Q^\aA(W)}{\pl W^\aB}\equiv0\,,
\end{align}
which can also we rewritten as $Q^2=0$ upon introducing the odd vector field
\begin{align}
Q= Q^\aA \frac{\overrightarrow{\pl} }{\pl W^\aA}\,.
\end{align}
An important consequence of the integrability is that the equations enjoy the gauge symmetries
\begin{align}
\delta W^\aA=d\xi^\aA+\xi^\aB\frac{\overrightarrow{\pl} Q^\aA(W)}{\pl W^\aB}\label{unfldgaugesymmetry}\,,
\end{align}
with the gauge parameters $\xi^{\aA}$ being differential forms of appropriate degrees. Now we would like to stop paying attention to the PDE's that the unfolded equations may encode and focus upon the consistency conditions \eqref{GrandJacobi}, a viewpoint pioneered already in \cite{Vasiliev:1988sa}. The passage to formal structures is performed by noticing that the fact that $W^\aA$ are differential forms can be relaxed to the requirement that they are coordinates on some graded manifold, so that each $W^\aA$ is assigned some non-negative degree. The de Rham differential $d=dx^\mm \frac{\pl}{\pl x^\mm}$, in its turn,  may also be understood as a homological vector field on the odd tangent bundle of the space-time manifold; in so doing, the differentials $dx^\mm$ are treated as odd coordinates in fibers.  If we now regard $d$ and $Q$ as  abstract homological vector fields on abstract graded manifolds, then the fields $W^\aA$ just provide the coordinate description of  smooth maps between the two $Q$-manifolds and the space of solutions to the field equations \eqref{dWQ}  is identified with the maps that relate the homological vector fields, i.e. $W^\ast(d)=Q$. Such a map may not exist for a given pair of $Q$-manifolds in which case the space of solutions is empty.\footnote{It is worth stressing that a formal unfolded system may not correspond to well-defined differential equations in space-time if $W^\aA$ is promoted to a field and $d$ is identified with $dx^\mm \frac{\pl}{\pl x^\mm}$. While all partial-differential equations can be in principle written in unfolded form, as we will comment also below, some of the formal unfolded equations are ill-defined as PDE's. The precise description of the class of unfolded equations that lead to well-defined PDE's is not yet known.
}

It is clear that all the information about any theory is encoded in the structure constants. Moreover, it is also not surprising that the structure constants bear certain algebraic meaning whenever some of the fields are gauge connections.

\subsection{Unfolding Higher-Spin Fields}
\label{sec:unfoldingHS}
A priory it may be unclear how to reformulate a given theory (set of differential equations) in the unfolded language, especially when no theory is available, but if it is known passage to unfolding can always be done \cite{Barnich:2010sw}. A good starting point in the higher-spin case is to rewrite the free equations of motion as unfolded equations, see the original work \cite{Vasiliev:1986td} for the $4d$ case. Free equations together with the knowledge of a higher-spin algebra \cite{Fradkin:1986ka,Vasiliev:1986qx} determine the field content and fix some of the boundary conditions to the non-linear deformations, which can then be systematically sought for. This is what we review below, the main difficulty being to explain why the unfolded equations do describe free fields of all spins. We begin with the Bargmann-Wigner equations, unfold them, and then attach Fronsdal fields via higher-spin connections.

\subsubsection{Free Data}
\label{sec:freedata}
Complementary to the Fronsdal fields, massless fields in $4d$ can be described by spin-tensors $C_{\ga_1...\ga_{2s}}$ and $C_{\gad_1...\gad_{2s}}$ of type $(2s,0)$ and $(0,2s)$, respectively, which are also known as higher-spin Weyl tensors. They obey the Bargmann-Wigner equations\footnote{$\ga,\gb,...=1,2$ and $\gad,\gbd,...=1,2$ are the indices of the fundamental and anti-fundamental representations of the Lorentz algebra $sl(2,\mathbb{C})\sim so(3,1)$. There are two invariant tensors, $\epsilon^{\ga\gb}$ and $\epsilon^{\gad\gbd}$, $\epsilon^{\ga\gb}=-\epsilon^{\gb\ga}$, $\epsilon^{12}=1$ and {\it  idem.} for $\epsilon^{\gad\gbd}$, which are used to raise and lower the indices according to $v^\ga=\epsilon^{\ga\gb} v_\gb$, $v^\ga\epsilon_{\ga\gb}=v_\gb$, where $\epsilon^{\ga\gb}\epsilon_{\ga\gc}=\delta^\gb_\gc$. As a part of the vector-spinor dictionary an $so(3,1)$-vector, say $v^a$, corresponds to a bi-spinor $v^{\ga\gad}$ via the usual $\sigma$-matrices, $\sigma^{\ga\gad}_m$. For example, $dx^\mm$ can be replaced by $dx^{\ga\gad}$ in flat space.}
\begin{align} \label{BW}
\epsilon^{\gb\gc}\frac{\pl}{\pl x^{\gb\gbd}} C_{\gc\ga_2...\ga_{2s}}&=0\,,  & 
\epsilon^{\gbd\gdd}\frac{\pl}{\pl x^{\gb\gbd}} C_{\gdd\gad_2...\gad_{2s}}&=0\,.
\end{align}
It is not difficult to see that the equations can be rewritten as an equation of unfolded form
\begin{align}
 dx^{\ga\gad}\left(\frac{\pl}{\pl x^{\ga\gad}} -\frac{\pl^2 }{\pl y^\ga \pl \bry^\gad}\right)C(y,\bry|x)&=0\,,
\end{align}
where the HS Weyl tensors together with all nontrivial on-shell derivatives thereof are packed into a generating function
\begin{align}
C(y,\bry|x)&= \sum_{m,n} \frac{1}{m!n!} C_{\ga_1...\ga_{m},\gad_1...\gad_{n}}\, y^{\ga_1} \ldots y^{\ga_m}\, \bry^{\gad_1} \ldots \bry^{\gad_n}\,.
\end{align}
These equations encode the Bargmann-Wigner equations for $C(y,0|x)$ and $C(0,\bry|x)$, which are generating functions for the fields in \eqref{BW}. The rest of the equations identifies other components with the on-shell derivatives of the HS Weyl tensors:
\begin{align}
C_{\ga_1...\ga_{2s+k},\gad_1...\gad_{k}}&\sim \frac{\pl}{\pl x^{\ga_1\gad_1}}\ldots\frac{\pl}{\pl x^{\ga_k\gad_k}} C_{\ga_{k+1}...\ga_{2s+k}}\,, && \text{{\it idem.} for }C_{\ga_1...\ga_{k},\gad_1...\gad_{2s+k}}\,.
\end{align}
The zeroth component $C(0,0|x)$ turns out to obey the massless Klein-Gordon equation. A small change of notation allows one to cast the equations into the unfolded form
\begin{align}
 d C(y,\bry|x)&= h^{\ga\gad}\frac{\pl^2 }{\pl y^\ga \pl \bry^\gad} C(y,\bry|x)\,, && dh^{\ga\gad}=0\,, \label{flatBW}
\end{align}
where we introduced the vielbein one-form $h^{\ga\gad}=\sigma^{\ga\gad}_\mm dx^\mm =dx^{\ga\gad}$. We added $dh=0$, so that $d$ of all the fields are expressed in terms of the other fields (or vanish). Therefore, the set of fields is $W^\aA=\{h^{\ga\gad},C_{\ga_1...\ga_{m},\gad_1...\gad_{n}}\}$. The next step would be to add spin-connection as to allow for coordinates other than Cartesian. Instead, we jump to $AdS_4$ that can be described as any non-degenerate solution of 
\begin{align} \label{adsdefinition}
   dh^{\ga\gad}&= \varpi\fud{\ga}{\gb}\wedge h^{\gb\gbd}+ \varpi\fud{\gad}{\gbd}\wedge h^{\ga\gbd}\,, & 
   \begin{aligned}
        d\varpi^{\ga\gb}&= \varpi\fud{\ga}{\gc} \wedge \varpi^{\gc\gb}+h\fud{\ga}{\gad}\wedge h^{\gb\gad}\,,\\
        d\varpi^{\gad\gbd}&= \varpi\fud{\gad}{\gdd} \wedge \varpi^{\gdd\gbd}+h\fdu{\ga}{\gad}\wedge h^{\ga\gbd}\,,
   \end{aligned}
\end{align}
where $\varpi^{\ga\gb}=\varpi^{\gb\ga}$ and $\varpi^{\gad\gbd}=\varpi^{\gbd\gad}$ are the (anti)-selfdual components of the spin-connection $\varpi^{a,b}_\mm dx^\mm$. Here the cosmological constant was chosen as to have $1$ in front of the $h\wedge h$ terms. The $AdS_4$ lift of the unfolded Bargmann-Wigner equations is \cite{Vasiliev:1988sa}
\begin{align}\label{DCeq}
\tadD C(y,\bry|x)&\equiv \nabla C+ih^{\ga\gad}(y_\ga\bry_\gad-\pl_\ga\pl_\gad)C=0\,,
\end{align}
where we rescaled the fields as to account for the $i$-factor in $\tadD$,
which is more natural from the HS algebra point of view. $\nabla$ is the Lorentz-covariant derivative:
\begin{align}
\nabla&=d-\varpi^{\ga\ga}y_\ga \pl_\ga-\varpi^{\gad\gad}\bry_\gad \pl_\gad\,,\label{LorentzDer}
\end{align}
which rotates in the right way all the spinorial indices that are contracted with $y^\ga$ and $\bry^\gad$. Upon restoring the cosmological constant $\Lambda$ it can be seen that the new, as compared to \eqref{flatBW}, term $ h^{\ga\gad} y_\ga\bry_\gad$ is of order $\Lambda$ and vanishes in the flat limit, where  in Cartesian coordinates, i.e. $\varpi^{\ga\gb},\varpi^{\gad\gbd}=0$, we find \eqref{flatBW}. 

The Fronsdal field $\Fron_{\mm_1...\mm_s}$ is represented by two spin-tensors $\phi_{\ga_1...\ga_s,\gad_1...\gad_s}$ and $\phi'_{\ga_1...\gad_{s-2},\gad_1...\gad_{s-2}}$ that correspond to the trace-free part of $\Fron_{\mm_1...\mm_s}$ and its trace. The Bargmann-Wigner fields, or the HS Weyl tensors, are the order-$s$ curls of the Fronsdal fields:
\begin{align} \label{weyldefinition}
C_{\ga_1...\ga_{2s}} &=\nabla\fdu{\ga_1}{\gad_1}...\nabla\fdu{\ga_s}{\gad_s} \phi_{\ga_{s+1}...\ga_{2s},\gad_1...\gad_s}\,, & 
C_{\gad_1...\gad_{2s}} &=\nabla\fud{\ga_1}{\gad_1}...\nabla\fud{\ga_s}{\gad_s} \phi_{\ga_1...\ga_s,\gad_{s+1}...\gad_{2s}}\,,
\end{align}
where the symmetrization over the $2s$ free indices is implied in both of the cases. The Weyl tensors can be shown to be gauge-invariant and to be consistent with \eqref{DCeq}. The same time the HS frame-like fields can be packed into a similar generating function $\omega=\omega_\mm(y,\bry|x)dx^\mm$. The Fronsdal fields reside in the diagonal components of $\omega$:
\begin{align}
\Phi_{\mm_1...\mm_s}&=\omega_{\mm_1|\ga_2...\ga_{s},\gad_2...\gad_{s}} h^{\ga_2\gad_2}_{\mm_2}\ldots h^{\ga_s\gad_s}_{\mm_s} +\text{permutations}\,.
\end{align}
The appropriate equations for $\omega$ read \cite{Vasiliev:1986td,Vasiliev:1988sa}:
\begin{align}\label{omegaeq}
&\adD \omega\equiv\nabla \omega-h^{\ga\gad}(y_\ga\pl_\gad+\bry_\gad\pl_\ga)\omega=\mathcal{V}(h,h,C)\,,
\end{align}
where the term on the right-hand side glues $C$ to the $\omega$-equations and is
\begin{align}\label{OMSTcocycle}
\mathcal{V}(h,h,C)&= h\fud{\ga}{\gnd}\wedge h^{\gb\gnd}\pl_\ga\pl_\gb C(y,\bry=0)+ h\fdu{\nu}{\gad}\wedge h^{\nu\gbd}\pl_\gad\pl_\gbd C(y=0,\bry)\,.
\end{align}
Eq.\eqref{omegaeq} sets to zero almost all components of $\adD \omega$. It can be shown that it imposes the Fronsdal equations, see e.g. \cite{Didenko:2014dwa}. The gluing term makes a dynamically trivial equation \eqref{weyldefinition} that identifies some components of $C$ with the order-$s$ curl of the Fronsdal field and should be read from right to left as a definition \eqref{weyldefinition}.

It is high time to introduce the higher-spin algebra $\mathfrak{hs}$ that explains the field content and most of the equations. We would like to study bosonic fields only and require the generating functions to have Taylor coefficients with even number of spinorial indices, i.e. $C(y,\bry)=C(-y,-\bry)$ and $\omega(y,\bry)=\omega(-y,-\bry)$. 

The relevant higher-spin algebra $\mathfrak{hs}$ is simply the even part of the Weyl algebra $A_2$ in four generators $\hat y_\ga, \hat \bry_\gad$, \cite{Vasiliev:1986qx}. The canonical normalization of the defining relations is 
\begin{align}
[\hat y_\ga,\hat y_\gb]&=2i\epsilon_{\ga\gb}\,, &
[\hat \bry_\gad,\hat \bry_\gbd]&=2i\epsilon_{\gad\gbd}\,.
\end{align}
The quadratic monomials can be shown to form the anti-de Sitter algebra $sp(4)\sim so(3,2)$ 
\begin{align}
P_{\ga\gad} &= -\frac{i}{4}\{\hat y_\ga,\hat \bry_\gad\}\,, & L_{\ga\gb}&=-\frac{i}{4} \{\hat y_\ga,\hat y_\gb\}\,,
& \bar{L}_{\gad\gbd}&=-\frac{i}{4} \{\hat \bry_\gad,\hat \bry_\gbd\}\,,
\end{align}
where $P$, $L$ and $\bar{L}$ are the generators of translations and Lorentz transformations $sl(2,\mathbb{C})\sim so(3,1)$. It is also convenient to replace operators with symbols thereof, which are functions of commutative variables $y^\ga$, $\bry^\gad$, and multiply symbols with the Moyal-Weyl star-product:
\begin{align}
(f\star g)(y,\bry) &= f(y,\bry)\exp i\left[\frac{\overleftarrow{\pl}}{\pl y^\ga}\epsilon^{\ga\gb}\frac{\overrightarrow{\pl}}{\pl y^\gb}+\frac{\overleftarrow{\pl}}{\pl \bry^\gad}\epsilon^{\gad\gbd}\frac{\overrightarrow{\pl}}{\pl \bry^\gbd}\right]g(y,\bry)\,.
\end{align}
The elements of the HS algebra, $\mathfrak{hs}$, are identified with even symbols $f(y,\bry)=f(-y,-\bry)$, which makes the bosonic projection. The background fields can be packed into a single $sp(4)\in \mathfrak{hs}$ connection
\begin{align}
\Omega&= \frac12 \varpi^{\ga\gb} L_{\ga\gb}+h^{\ga\gad} P_{\ga\gad}+\frac12 \varpi^{\gad\gbd} \bar{L}_{\gad\gbd}
\end{align}
and the defining equations \eqref{adsdefinition} of $AdS_4$ can be recognized as a flat $sp(4)$-connection, $d\Omega=\Omega\star \Omega$. The covariant derivatives \eqref{DCeq}, \eqref{omegaeq} turn out to be
\begin{align}
\adD \omega&= d\omega-\Omega\star \omega+\omega\star \Omega\,, &
\tadD C&=dC -\Omega\star C+C\star \pi(\Omega)\,,
\end{align}
where one of the most important points is that $\tadD$ is a twisted derivative, where $\pi$ flips the sign of the translation generators, $\pi(P_{\ga\gad})=-P_{\ga\gad}$ and leaves the Lorentz generators intact. It is realized as the HS algebra automorphism: $\pi(f)(y,\bry)=f(-y,\bry)=f(y,-\bry)$.

The term $\mathcal{V}(\Omega,\Omega,C)=\mathcal{V}(h,h,C)$ will be studied in the next Section. It is responsible for the identification \eqref{weyldefinition}, which looks like an empty equation. Nevertheless, there is a sense in which $\mathcal{V}(\Omega,\Omega,C)$ is a true cohomology: no redefinition $\omega\rightarrow \omega +f(\Omega,C)$ can trivialize it. If it were the case the connection $\omega$ would be pure gauge, $\adD \omega=0$.

It is worth stressing that the free equations are not only formal, but are well-defined differential equations, whose content is to impose the Fronsdal equations on fields with $s=0,1,2,3,...$ and express the other components of $\omega$ and $C$ as derivatives of the Fronsdal fields. We also note that $C$ contains derivatives of the Fronsdal fields of any order.

\subsubsection{Nonlinear Unfolded Equations}
\label{sec:nonlinearunfolding}
The main consequence of having unfolded equations for free higher-spin fields is the identification of the field content: the one-form gauge connection $\omega$ of the HS algebra $\mathfrak{hs}$ and the zero-form $C$ in the peculiar twisted-adjoint representation, so that the coordinates of the $Q$-manifold are $W^\aA=\{\omega,C\}$. Therefore, the full unfolded equations should read \cite{Vasiliev:1988sa}
\besubeqs
\begin{align}
&d\omega=Q^{\omega}(\omega,C)\label{xpaceseqXA}\,,\\
&dC=Q^C(\omega, C)\,,\label{xpaceseqXB}
\end{align}
\esubeqs
where $Q^\omega$ and $Q^C$ are two structure functions related by the nilpotency of $Q=Q^\omega \pl_\omega +Q^C\pl_C$. It would be hard to guess the full system at once and a natural expansion scheme is to treat $C$ as an expansion parameter (in fact, infinitely many of such parameters):
\besubeqs
\label{Cexpansion}
\begin{align}
&Q^{\omega}(\omega,C)=\mathcal{V}(\omega,\omega)+\mathcal{V}(\omega,\omega,C)+
\mathcal{V}(\omega,\omega,C,C)+...\,,\\
&Q^C(\omega, C)=\mathcal{U}(\omega,C)+\mathcal{U}(\omega,C,C)+\mathcal{U}(\omega,C,C,C)+...\,.
\end{align}
\esubeqs
The knowledge of the free equations suggests that the first two vertices are\footnote{The $\pi$-automorphism can be extended from the $AdS$-subalgebra of the HS algebra to the full HS algebra since any HS algebra results from the universal enveloping algebra of the AdS algebra. In the $4d$ case, it is obvious that $\pi(f)=f(-y,\bry)=f(y,-\bry)$ works for the whole HS algebra.} 
\begin{align}\label{trivialcocycles}
\mathcal{V}(\omega,\omega)&=\omega\star\omega\,,
&\mathcal{U}(\omega, C)&=\omega\star C-C\star \pi(\omega)\,.
\end{align}
The gluing term $\mathcal{V}(\Omega,\Omega,C)$, where $\Omega$ is a flat $sp(4)$-connection, should be the $AdS_4$ limit of some $\mathcal{V}(\omega,\omega,C)$ that is defined on the full HS algebra $\mathfrak{hs}$. Indeed, if we replace $\omega\rightarrow \Omega+\omega$ and pick the terms of the zeroth and of the first order in $\omega$ we find exactly the equations \eqref{adsdefinition}, \eqref{DCeq}, \eqref{omegaeq} from the previous Section, provided that $\mathcal{V}(\Omega,\Omega,C)$ gets reduced to \eqref{OMSTcocycle}.

Therefore, the expansion parameter $C$ is the deviation of $\omega$ from a flat connection. Flat connection is a topological solution since locally it is pure gauge. It should be stressed that this expansion scheme is in some sense orthogonal to the usual weak-field expansion where the vacuum solution is $AdS$ and it is the Fronsdal fields that are treated as small. As an illustration let us write the first two orders in the weak-field expansion scheme, see \cite{Boulanger:2015ova} for explicit results, 
\begin{align}
\omega&=\Omega + \omegaone+\omegatwo+...\,, && C=0+\Cone+\Ctwo+...\,,
\end{align}
and confront it with the scheme above:
\besubeqs\label{xpaceseqBQQ}
\begin{align}
&\adD \omegaone=\mathcal{V}(\Omega,\Omega,\Cone)\,, \\&\tadD \Cone=0\,,\\
&\adD \omegatwo-\mathcal{V}(\Omega,\Omega,\Ctwo)=\omegaone\star\omegaone+
\mathcal{V}(\Omega,\omegaone,\Cone)+
\mathcal{V}(\Omega,\Omega,\Cone,\Cone)\label{xpaceseqBA}\,,\\
&\tadD \Ctwo=\omegaone\star \Cone-\Cone\star \pi(\omegaone)+\mathcal{V}(\Omega,\Cone,\Cone)\label{xpaceseqBB}\,.
\end{align}
\esubeqs
Here $\omegaone$, $\Cone$ are free fields and the equations are equivalent to the Fronsdal ones, while the second-order fields $\omegatwo$, $\Ctwo$ have sources bilinear in $\omegaone$, $\Cone$. It is clear that the pure star-product, whose only trace in the free equations is to build up the $AdS$ covariant derivative, contributes to the interactions at the second order and higher via $\omegaone\star\omegaone$. Likewise, the term $\mathcal{V}(\Omega,\Omega,\Cone)$ at the free level serves to identify the order-$s$ derivative of the Fronsdal field with the HS Weyl tensor, but its HS algebra covariantization $\mathcal{V}(\omega,\omega,C)$ contributes to the second and higher order interactions. The term $\mathcal{V}(\Omega,\Omega,\Cone,\Cone)$ should have an interpretation of the gauge-invariant part of the HS stress-tensors and its full structure, $\mathcal{V}(\omega,\omega,C,C)$, starts to be effective at the fourth order in weak fields, but at the second order in $C$.

The conclusion is that the expansion in scheme where $C$ is treated small is more powerful than the usual weak-field expansion over $AdS$ since it keeps track of the full HS algebra covariance and resums the terms that would contribute to different orders within the weak-field expansion.

\section{Unfolding and Hochschild Cohomology}
\label{sec:UnfldHoch}
The upshot of the review above is that the unfolded equations for HS fields should be looked for in the following form\footnote{At the beginning we follow the seminal paper \cite{Vasiliev:1988sa}, but the interpretation of some of the steps is somewhat new. } 
\besubeqs
\begin{align}
&d\omega=\omega\star\omega+\mathcal{V}(\omega,\omega,C)+...\,,\\
&dC=\omega\star C-C\star \pi(\omega)+...\,,
\end{align}
\esubeqs
where the first vertex to be determined is $\mathcal{V}(\omega,\omega,C)$.\footnote{The linearized and more generally weak-field analysis reveals that this vertex should contain two parts: one of them makes the right-hand side of the Fronsdal equations, i.e. it is zero in the free case and contains some interaction terms that are nonlinear in the Fronsdal fields otherwise. Another part is the nonlinear completion of the HS Weyl tensor definition \eqref{weyldefinition} that is compatible with the interaction terms. Both of them have to be HS algebra covariant.} There is also a boundary condition \eqref{OMSTcocycle} that it should obey, which comes from the free equations over $AdS$. The Frobenius integrability $dd=0$ or $QQ=0$ implies the following consistency condition
\begin{align}
\begin{aligned}
&\mathcal{V}(\omega\star\omega,\omega,C)-\mathcal{V}(\omega,\omega\star \omega,C)+\mathcal{V}(\omega,\omega,\omega\star C-C\star \pi(\omega))+\\
&\qquad\qquad-\omega\star\mathcal{V}(\omega,\omega,C)+\mathcal{V}(\omega,\omega,C)\star \omega=0\,,
\end{aligned}
\end{align}
which should be understood as a cohomology problem modulo trivial vertices induced by nonlinear field-redefinitions:
\begin{align}
\omega\rightarrow\omega +g(\omega,C)\,.
\end{align}
Formally, the problem is that of the Chevalley-Eilenberg cohomology of the HS algebra $\mathfrak{hs}$, viewed as a Lie algebra, valued in $\mathfrak{hs}$ and with coefficients in the twisted-adjoint representation, which is difficult to say anything about. Fortunately, there are $AdS/CFT$-inspired simplifying assumptions, which were present already in \cite{Vasiliev:1988sa} for a different reason. We would like to recall that higher-spin algebras are associative, while in the equations above we seem to consider them as Lie algebras constructed via commutator.

\paragraph{Gauging CFT Global Symmetry in the Bulk and Morita Invariance.} It turns out that one can considerably simplify the problem by noticing that the very possibility to have global symmetries in the CFT duals of higher-spin theories allows one to replace the Chevalley-Eilenberg cohomology with the Hochschild one.

Bearing AdS/CFT in mind we expect HS theories to be generically duals of free CFT's. In any free CFT it should be possible to add global symmetries, say $u(M)$.\footnote{ Here we discuss the global symmetries that remain after the singlet constraint is imposed. } In the dual picture of HS theories global symmetries on the CFT side should result in a local gauge group, say $u(M)$. Effectively, this means that each HS field is now `matrix-valued'. Mathematically, the possibility to extend free CFT's with global symmetries results in replacing the original HS algebra $\mathfrak{hs}$ with the tensor product\footnote{This corresponds to the simplest $u(M)$-gauging. For $so(M)$ or $usp(M)$ gaugings there are certain symmetry constraints, e.g. fields with odd spins should be in adjoint of $so(M)$ and fields with even spins be symmetric matrices of $so(M)$, see e.g. \cite{Konstein:1989ij}. The simplest $u(M)$ option will suffice for our purpose.} $\mathfrak{hs}\otimes \mathrm{Mat}_M$. Therefore, one should study the Chevalley-Eilenberg cohomology of $\mathfrak{hs}\otimes \mathrm{Mat}_M$ viewed as a Lie algebra.

Firstly, there is a fairly general statement that the Chevalley-Eilenberg cohomology of a Lie algebra obtained by tensoring an associative algebra, say $\mathfrak{hs}$, by a matrix algebra is essentially equivalent to the Hochschild cohomology of  $\mathfrak{hs}$, provided that the size of matrices is large enough.\footnote{We would like to note that the condition of `size large enough' is important.  For smaller matrices, there can be additional solutions that we miss. Such solutions may, for example, be relevant for super-symmetric extensions of higher-spin theories that, as in the case of super-gravities, rely on non-trivial low-dimensional Fierz identities. So far all the super-symmetric extensions of higher-spin theories were obtained by tensoring with Clifford algebras \cite{Konshtein:1988yg,Sezgin:2012ag}. } The second step is to note that due to the Morita invariance the Hochschild cohomology of any associative algebra, say $\mathfrak{hs}$, tensored with matrices of any size is isomorphic to the Hochschild cohomology of $\mathfrak{hs}$. These facts strengthen again the idea that HS theories should be based on associative  algebra structures. 

In practice, the advantage is that we can split all the functionals according to the ordering $\omega$ and $C$, thinking of them as having additional matrix factors. Essentially following \cite{Vasiliev:1988sa} let us write down the consistency relations for $\mathcal{V}$ that include all possible orderings:
\begin{align}
\mathcal{V}&=\mathcal{V}_1(\omega,\omega,C)+\mathcal{V}_2(\omega,C,\omega)+\mathcal{V}_3(C,\omega,\omega)\,.
\end{align}
Denoting $\tilde{\omega}\equiv\pi(\omega)$, the consistency relations for different orderings are:
\besubeqs
\begin{align}
0&=\mathcal{V}_1(\omega\star \omega,\omega,C)-\omega\star \mathcal{V}_1(\omega,\omega,C)+\mathcal{V}_1(\omega,\omega,\omega\star C)-\mathcal{V}_1(\omega,\omega\star \omega,C)\,,\\
0&=\mathcal{V}_1(\omega,\omega,C)\star \omega-\mathcal{V}_1(\omega,\omega,C\star \tilde\omega)+\mathcal{V}_2(\omega\star \omega,C,\omega)-\omega\star \mathcal{V}_2(\omega,C,\omega)-\mathcal{V}_2(\omega,\omega\star C,\omega)\,,\\
0&=\mathcal{V}_2(\omega,C,\omega)\star\omega-\mathcal{V}_2(\omega,C,\omega\star \omega)-\omega\star \mathcal{V}_3(C,\omega,\omega)+\mathcal{V}_3(\omega\star C,\omega,\omega)+\mathcal{V}_2(\omega,C\star \tilde\omega,\omega)\,,\\
0&=\mathcal{V}_3(C,\omega,\omega)\star \omega-\mathcal{V}_3(C,\omega,\omega\star \omega)+\mathcal{V}_3(C,\omega\star \omega,\omega)-\mathcal{V}_3(C\star \tilde\omega,\omega,\omega)\,,
\end{align}
\esubeqs
and likewise for the redefinitions we have
\begin{align}
\omega\rightarrow\omega +g_1(\omega,C)+g_2(C,\omega)\,.
\end{align}
In the last step we can look for a solution of the system with $\mathcal{V}_2=\mathcal{V}_3=0$ as it is equivalent to the solution of the full system, i.e. by reducing everything to the Hochschild cohomology of the HS algebra $\mathfrak{hs}$:
\besubeqs\label{twocons}
\begin{align}
0&=\mathcal{V}_1(\omega\star \omega,\omega,C)-\omega\star \mathcal{V}_1(\omega,\omega,C)+\mathcal{V}_1(\omega,\omega,\omega\star C)-\mathcal{V}_1(\omega,\omega\star \omega,C)\,,\\
0&=\mathcal{V}_1(\omega,\omega,C)\star \omega-\mathcal{V}_1(\omega,\omega,C\star \tilde\omega)\,.
\end{align}
\esubeqs
The discussion in this Section is dimensional independent and the same arguments apply to any HS theory in arbitrary number of space-time dimensions.\footnote{This is true unless we would like to have mixed-symmetry gauge fields that are described by forms of higher degrees \cite{Alkalaev:2003qv,Boulanger:2008up,Skvortsov:2009zu}.}

\subsection{Hochschild Cocycle}
\label{sec:HochCocycle}
Bearing in mind that one can take advantage of associative structures we can rewrite the consistency conditions \eqref{twocons} as
\besubeqs
\begin{align}
&\mathcal{V}(a\star b,c,d)-\mathcal{V}(a,b\star c,d)+\mathcal{V}(a,b,c\star d)-a\star\mathcal{V}(b,c,d)=0\,,\\
&\mathcal{V}(a,b,c)\star d=\mathcal{V}(a,b,c\star \tilde{d})\,,\label{Equivariance}
\end{align}
\esubeqs
where we use $a,b,c,d$ for four arbitrary elements of the HS algebra. The last equation, which we refer to as {\it equivariance condition}, allows one to solve $\mathcal{V}(a,b,c)$ for the third argument:
\begin{align}
  \mathcal{V}(a,b,c)=\Phi(a,b)\star \tilde{c}  \,.
\end{align}
As a result, the first equation is of the form $E\star \tilde{d}=0$, where 
\begin{align}\label{sptwoHochschildA}
    E=- a\star \Phi(b,c)+\Phi(a\star b,c)-\Phi(a,b\star c) +\Phi(a,b)\star \tilde{c}=0\,.
\end{align}
This is one of the most important equations. It implies that the first order deformation is governed by the Hochschild two-cocycle $\Phi(a,b)$ with values in the twisted-adjoint representation of the higher-spin algebra. Therefore, the vertex is $\mathcal{V}(\omega,\omega,C)=\Phi(\omega,\omega)\star \tilde{C}$.

We would like to massage it a little bit assuming that there is a non-degenerate super-trace operation $\str$ on the HS algebra such that $\str(x\star y)=\str (\tilde{y}\star x)$. Then $E=0$ is obviously equivalent to $\str (E\star \tilde{d})=0$ for arbitrary $d$. On the other hand, we can write
\begin{align}
   \mathrm{str} (E\star \tilde{d})=\mathrm{str}\left(- a\star \Phi(b,c)\star \tilde{d}+\Phi(a\star b,c)\star \tilde{d}-\Phi(a,b\star c)\star \tilde{d} +\Phi(a,b)\star \tilde{c}\star \tilde{d}\right)=0
\end{align}
and use the property of the super-trace $str(x\star y)=str (\tilde{y}\star x)$ as to move $c,d$ to the left:
\begin{align}
   \mathrm{str }(E\star \tilde{d})=\mathrm{str}\left(- d\star a\star \Phi(b,c)+d\star\Phi(a\star b,c)-d\star\Phi(a,b\star c) +c\star d\star\Phi(a,b)\right)=0\,.
\end{align}
Again, the equation above is the equation for the Hochschild cocycle
\begin{align}
    f(x\star y| z,w)-f(x|y\star z, w) +f(x|y,z\star w)-f(w\star x|y,z)=0\,,
\end{align}
where the cocycle is
\begin{align}
    f(x|y,z)&= \mathrm{str}(x\star \Phi(y,z))\,.
\end{align}
It is clear that $f(x|y,z)$ is a two-cocycle of the HS algebra $\mathfrak{hs}$ with values in $\mathfrak{hs}^*$. Canonically, values in $\mathfrak{hs}^*$ can be traded for a functional on $\mathfrak{hs}$, which is what we did here-above. This is the case whenever a higher-spin algebra is the Weyl algebra and the $\pi$-automorphism is realized as $\tilde f(y)=f(-y)$, e.g. in the $4d$ bosonic HS theory.  Assuming the natural $\mathbb{Z}_2$-grading on the Weyl algebra, there exists a super-trace $\mathrm{str}\, f(y)=f(0)$, see e.g. \cite{Vasiliev:1999ba,Pinczon2005}.

Therefore, we see that the first-order deformation of the higher-spin equations is governed by the Hochschild cocycle of the higher-spin algebra. In practice, it is also convenient to work with \eqref{sptwoHochschildA}. Let us emphasize again that the knowledge of this single cocycle gives information about all orders in the usual weak-field expansion scheme. The obstructions to promoting this vertex to higher orders in $C$'s are controlled by the third cohomology groups of a particular HS algebra. For  the $4d$ HS algebra (or more generally for HS algebras that are Weyl algebras) all the obstructing groups seem to vanish (see Appendix \ref{app:Weyl}). This is also confirmed by the results of \cite{Vasiliev:1989yr}, where the second-order order deformation was explicitly constructed and by the Vasiliev equations \cite{Vasiliev:1990vu,Vasiliev:1999ba}. 
It is likely the case for all of the HS algebras, so that the higher-order interactions are unobstructed and simply provide a nonlinear completion of $V(\omega,\omega,C)$.\footnote{\label{higherorders}Going to higher orders, at every second order one can use $\Phi(\omega,\omega)\star (C\star \tilde{C})^k\star \tilde{C}$ as a new interaction vertex (there are two such vertices, which is what makes it nontrivial, otherwise it can be obtained via $C\rightarrow C+(\tilde{C}\star C)^k\star C$ redefinition), which is exactly the ambiguity in one function that the original Vasiliev equations have \cite{Vasiliev:1992av}. This ambiguity seemingly leads to infinitely many couplings constants that start to affect higher correlators on the CFT side while having no effect on the three-point functions, which is inconsistent with CFT,  \cite{Maldacena:2012sf}. Such vertices can be shown to be too non-local when one is trying to pass from formal unfolded equations to PDE's \cite{Boulanger:2015ova}. Not surprisingly, they destroy the near boundary analysis too \cite{Vasiliev:2015mka}. Therefore, such vertices should be forbidden. This is one of the simplest examples of formal structures that make no sense within field theory, AdS/CFT or just as PDE's.  }

Additional simplifications take place in the case of the $4d$ HS theory. The simplest bosonic HS algebra $\mathfrak{hs}$ is the even subalgebra of the Weyl algebra $A_2$ with $y_\ga$ and $\bry_\gad$ as generators. It turns out that the problem can be reduced to just $A_1$ of either $y_\ga$ or $\bry_\gad$. Indeed, explicit realization of the $\pi$-automorphism is $\pi(f)(y,\bry)=f(-y,\bry)=f(y,-\bry)$, where the last equality is thanks to the truncation to the even subalgebra. Moreover, the Lorentz symmetry does not allow us to mix $y_\ga$ and $\bry_\gad$. Therefore, one should look for the deformation $\Phi(a,b)$ that acts either on $y$ or $\bry$. We have presented some plausible arguments, but this fact can be rigorously proved, see Appendix \ref{app:Weyl}, and, in fact, the cohomology is known. As a result, one has to solve \eqref{sptwoHochschildA} for arguments in $A_1$, i.e. for the functions of $y$ only. We note that $f(-y,\bry)=f(y,-\bry)$ but as functions of only $y$ (or $\bry$) they are not constrained by parity. Therefore, the first vertex in the $4d$ theory is determined by the Hochschild cocycle of $A_1$. We will give explicit formulas below when it comes to the general case of  $A_{n}$.

As a historical comment, we would like to note that the existence of the Hochschild cocycle discussed above, as well as of its higher-dimensional generalizations, has been long known. Its analytical structure was discussed e.g. in \cite{FeiginTsygan}. It turns out that the Hochschild homology is much simpler than the cohomology and from the fact that there is a natural pairing of cycles with cocycles it follows that there exists a nontrivial Hochschild cocycle. Its explicit form, however, was not available before \cite{FFS} and for the simplest case of $A_1$ it was implicitly found in \cite{Vasiliev:1988sa}.

\section{Hochschild Cohomology and Formality}
\label{sec:HochFormality}
We have pointed out that the Hochschild cocycle of the relevant higher-spin algebra turns out to govern the deformations of the unfolded equations to the first order. The problem is how to construct such a cocycle. It turns out that there is an explicit formula for the cocycle \cite{FFS}, which results from the formality theorems \cite{Kontsevich:1997vb,Tsygan,Shoikhet:2000gw}. 

Most of this Section is devoted to the Hochschild cocycle for the general case of the Weyl algebra $A_{n}$. Its finite-dimensional subalgebra is $sp(2n)$ which, for different $n$, has already showed up in the HS studies \cite{Fronsdal:1985pd,Bandos:1998vz,Vasiliev:2001dc,Plyushchay:2003gv,Bandos:2005mb,Vasiliev:2007yc} and we also note that $sp(2,2)$ is the Lorentz algebra in $AdS_5$. We review the results of \cite{FFS} and extend them by using some of the ideas of \cite{Vasiliev:1989xz}. In particular, we show that the Hochschild cocycle is a transform of a remarkable function of $2n+1$ symplectic vectors that form a simplex and the cocycle condition has a geometric interpretation of cutting a polytope into simplices. This allows to represent the Hochschild cocycle as a coboundary in a larger space and explains the doubling trick behind the Vasiliev equations \cite{Vasiliev:1990vu}.

At the end we also sketch the idea of the Shoikhet-Tsygan formality \cite{Tsygan} that gives explicit formulas for the Hochschild cocycles. This part makes use of the formality nomenclature.

\subsection{Hochschild Cocycle of Weyl Algebra}
\label{sec:FFS}

Paying tribute to the conventions widely used in the higher-spin literature we define Weyl algebra $A_{n}$ with $2n$ generators as 
\begin{align}
[y_\ga,y_\gb]_\star &= 2iC_{\ga\gb}\,, &&\ga,\gb,...=1,...,2n\,,
\end{align}
where we immediately assume that we work with the symbols of the elements of the Weyl algebra and some ordering for $y_{\ga_1}...y_{\ga_k}$ is chosen. We prefer to choose the Weyl ordering, i.e. the symbols correspond to totally symmetrized monomials $y_{\ga_1}...y_{\ga_k}$. Therefore, the product on the Weyl algebra is mapped to the Moyal-Weyl star-product 
\begin{align}
(f\star g)(y) &= f(y)\exp i\left[\frac{\overleftarrow{\pl}}{\pl y^\ga}C^{\ga\gb}\frac{\overrightarrow{\pl}}{\pl y^\gb}\right]g(y)\,.
\end{align}
In practice we have to work with multi-linear operators and for that reason it is convenient to use the form of the star-product adopted to several arguments:
\begin{align}
\begin{aligned}
(f\star g)(y) &= \exp i[-iy^\nu(\pl_1+\pl_2)_\nu +(\pl_1)_\nu (\pl_2)^\nu]f(y_1)g(y_2)\Big|_{y_i=0}=\\
&=\exp i [p_0\cdot p_1+p_0\cdot p_2 +p_1\cdot p_2] f(y_1)g(y_2)\Big|_{y_i=0}\,,
\end{aligned}
\end{align}
where in the last line we denoted $iy=p_0$, $\pl_1=p_1$, etc. The scalar product is as usual $q\cdot p\equiv q_\ga p_\gb C^{\ga\gb}$. We will omit $|_{y_i=0}$ hereafter. More information can be found in Appendix \ref{app:symbols}. For example, the star-product of several functions corresponds to
\begin{align}
(f_1\star...\star f_k)(y)&=\exp i\left[\sum_{0=i<j=k}p_i \cdot p_j\right] f_1(y_1)...f_k(y_k)\,.
\end{align}
The Hochschild cohomology is one-dimensional and is concentrated in degree $2n$, see Appendix \ref{app:Hochschild}. Therefore, the equation for the $A_{n}$ Hochschild cocycle, which is a natural generalization of the $A_1$-case \eqref{sptwoHochschildA}, reads
\begin{align}\label{HochCocycle}
   - a_1\star \Phi(a_2,...,a_{2n},a_{2n+1})+\Phi(a_1\star a_2,...,a_{2n},a_{2n+1})-... +\Phi(a_1,a_2,...,a_{2n})\star \tilde{a}_{2n+1}=0\,.
\end{align}
In the language of symbols of operators one has to find a differential operator $\hat\Phi(p_0,...,p_{2n})$ that obeys the analog of the equation above and is nontrivial, i.e. cannot be obtained as a coboundary. The case of $A_1$ can be approached by elementary methods, see Appendix \ref{app:lame} for more detail.

The solution for any $n$ was found by Feigin, Felder and Shoikhet (FFS) \cite{FFS} and has a remarkably simple form as a generating function of $p_i$:
\begin{align}\label{ffscocycle}
    \hat\Phi(p_0,...,p_{2n})&= \det |p_1,\ldots,p_{2n}|\int_\Sigma d^{2n}u\, \exp{i\left[\sum_{0\leq i<j\leq 2n} (1+2u_i-2u_j)(p_i\cdot p_j)\right]}\,,
\end{align}
where the integration is over the $2n$-simplex $\Sigma$:
\begin{align}
\Sigma&: && u_0=0\leq u_1 \leq u_2\leq \ldots\leq u_{2n}\leq1\,.\label{simplexA}
\end{align}
In addition to the cocycle property and manifest $sp(2n)$-invariance, which follows from the fact that $p_i\cdot p_j$ are contracted with the $sp(2n)$-invariant tensor $C^{\ga\gb}$, the FFS cocycle has a remarkable property that it is $sp(2n)$-basic:
\begin{align}\label{spbasic}
\sum_{i=1}^{i=2n} \Phi(f_0,...,f_{i-1},L_{\ga\gb},f_{i+1}...,f_{2n-1})(-)^i&=0\,, & L_{\ga\gb}&=-\frac{i}{4}\{y_\ga,y_\gb\}\,,
\end{align}
i.e. it vanishes whenever one of the arguments belongs to $sp(2n)$ that is generated by $L_{\ga\gb}$ and is anti-symmetrized with $f_1,\ldots, f_{2n-1}$. This property is advantageous in higher-spin theories whenever $sp(2n)$ is the (generalized) Lorentz symmetry. It implies that  the spin-connection does not appear outside the Lorentz-covariant derivative, which is a form of the equivalence principle.

\subsection{Geometric Interpretation}
\label{sec:Delta}
The FFS cocycle can be represented in the form
\begin{align}
    \int d^{2n}u\,\exp{i\left[ \sum_{0\leq i<j\leq 2n} p_i\cdot p_j+2u\cdot \sum_0^{2n} p_i\right]}\Delta(u+p_0,u+p_0+p_1,...,u+p_0+...+p_{2n})\,,
\end{align}
where $\Delta$ is a function of $2n+1$ arguments and it checks roughly speaking if a $(2n+1)$-tuple of vectors in $2n$-dimensional space forms a simplex such that it contains the origin.\footnote{In the case of $A_1$ similar representation was discovered in \cite{Vasiliev:1989xz}.} The precise definition is
\begin{align}
    \Delta(a_1,...,a_{2n+1})&=\int d\beta\, \sum_{i=1}^{2n+1}(-)^{i+1} \det |a_1,...,\hat{a}_i,...,a_{2n+1}| \delta^{2n}\left(\sum_{i=1}^{2n+1} \beta_i a_i\right)\delta \left(\sum \beta_j -1\right)\,,
\end{align}
where $\beta_j\geq0$. The prefactors are the volumes of certain simplices. The delta function with $\sum \beta_i a_i$ contributes only if the origin belongs to the simplex. It is straightforward to see that 
\begin{itemize}
    \item $\Delta$ is the characteristic function of the oriented simplex and takes three values $0,\pm1$;
    \item $\Delta$ is $Sp(2n)$ invariant, i.e. any linear symplectic map $a_i\rightarrow A a_i$ leaves it invariant;
    \item it is totally anti-symmetric under the permutation of all the arguments, while the FFS cocycle has more complicated symmetries: any permutation of the last $2n$ arguments needs to be accompanied with the appropriate rearrangement of the simplex;
    \item any linear transformation $A$ of all the vectors multiplies $\Delta$ by $\sign \det A $;
    \item $\Delta$ vanishes whenever two arguments coincide, which, together with other properties, implies that the FFS cocycle vanishes on $sp(2n)$, \eqref{spbasic};
    \item the Hochschild cocycle condition can be interpreted in the following geometric way:
    \begin{align}\label{hochdelta}
        \sum_i (-)^i\Delta(a_1,...,\hat{a}_i,...,a_{2n+2})=0
    \end{align}
    which expresses the fact that the polytope made of $2n+2$ points in $2n$-dimensional space can be split into $2n+2$ simplices. The origin then can be shown to belong to an even number (possibly zero) of such simplices with the appropriate sign factor and orientation ensuring the cancellation. 
    \item $\Delta$ is the Alexander–Spanier cocycle of $\mathbb{R}^{2n}$. Introducing the Alexander–Spanier differential $\pl$,
    \begin{align}
    (\pl f)(a_0,...,a_{k+1})&=\sum_{i=0}^{i=k+1} f(a_0,...,\hat{a}_i,...,a_{k+1})(-)^i\,,
    \end{align}
    we see that \eqref{hochdelta} amounts to $\pl \Delta=0$.
\end{itemize}
One of the crucial observations made in the case of $A_1$ in \cite{Vasiliev:1989xz} was that the cocycle property of $\Delta$ implies that it can formally be represented as its own coboundary:
\begin{align}
\Delta(a_1,...,a_{2n+1})&= (\pl \Delta)(z,a_1,...,a_{2n+1})\,,
\end{align}
where $z$ is a fixed $2n$ dimensional vector and it is not acted on by the differential $\pl$. There seems to be a contradiction with the fact that $\Delta$ corresponds to the cohomology. However, it is easy to see that the resulting coboundary (or the field-redefinition) is singular --- there is a singularity whenever any of the arguments coincides with $z$. There is a trick that still allows one to take advantage of the `coboundary' representation: $z$ can be thought of as a new variable that $y$ can never coincide with. This is essentially the rationale behind the doubling of oscillators that occur in the Vasiliev equations \cite{Vasiliev:1990vu}. We will make this more precise in the next Section.

\subsection{Vasiliev Double}
\label{sec:Double}
Given the FFS cocycle, let us rewrite the integral over the simplex as an integral over the hypercube, which can be constructed by blowing up some of the edges into faces --- the idea is to disentangle the variables that are constrained by inequalities:
\begin{align}
u_0=0\leq u_1 \leq u_2\leq \ldots\leq u_{2n}\leq1\,.
\end{align}
The appropriate change of variables is
\begin{align}
u_1&=t_0...t_{2n-1}\,, &u_2&=t_0...t_{2n-2}\,, &&\ldots &u_{2n-1}&=t_0t_1\,, &u_{2n}&=t_0\,,
\end{align}
with the Jacobian being $(t_0)^{2n-1}(t_1)^{2n-2}\ldots t_{2n-2}$. The variables in the FFS cocycle do not factorize, of course, and we cannot represent it as a star-product of several elements that depend on less variables. Nevertheless, the idea that a simplex can be replaced by a hypercube suggests that one can create the integrals and the integrands step by step: adding successively one dimension to the integration domain. 

To be precise, let us enlarge the Weyl algebra $A_{n}$ generated by $y_\ga$ by introducing auxiliary variables $z_\ga$ and postulate that for symbols of operators we have
\begin{align}
f(y)\star V(iy,iz;p_2,...,p_k)\rightarrow V(p_0+p_1,iz-2p_1;p_2,...,p_k) e^{i p_0\cdot p_1}\,. \label{newstar}
\end{align}
This induces some `entanglement' between $y$ and $z$, while the star-product on the functions of $y$ remains unchanged. This does not fix the star-product uniquely. It is convenient to assume that $y_\ga$ and $z_\ga$ form doubled Weyl algebra $A_{2n}$ with a star-product realization constrained by \eqref{newstar} --- we call it the Vasiliev double. Let also $\Gamma_n$ be an operator that resembles the contracting homotopy of the de Rham complex:
\begin{align}
\Gamma_n[f(z)]&=\int_0^1 dt\,t^n\, f(zt)\,.
\end{align}
The FFS cocycle can be created by repeating a number of simple steps. One starts with
\begin{align}
\varkappa&= \exp i[p_0\cdot (iz)]\,,
\end{align}
and it is easy to see that 
\begin{align}\label{FFSstepbystep}
\Phi(f_1,...,f_{2n})&= \det(p_1,...,p_{2n}) f_1(y)\star \Gamma_0\left[ f_2(y)\star \Gamma_1 \left[...f_{2n}(y)\star \Gamma_{2n-1} [\varkappa]\right]\right]\Big|_{z=0}\,.
\end{align}
At each of the steps $i=0,1,...,2n-1$ one multiplies the function from the left by the $(2n-i)$-th argument of $\Phi$; then one rescales $z\rightarrow z t_{2n-i+1}$ and multiplies by the Jacobian $t_{2n-i+1}^{2n-i}$. The determinant in front of the cocycle can also be created in this process if one starts with 
\begin{align}
\Zf&=\epsilon_{\ga_1...\ga_{2n}} \exp i[p_0\cdot (iz)]\,,
\end{align}
viewed as a $2n$-form in the $z$-space. Then $\Gamma$ becomes the contracting homotopy $\delta^{-1}$ of the de Rham complex in the $z$-space: given a $k$-form $f_{\ga_1...\ga_k}(z)$ we set 
\begin{align}
\delta^{-1}[f(z)]&=z^\nu\int_0^1 dt\,t^{k-1}\, f_{\nu\ga_1...\ga_{k-1}}(zt)\,.
\end{align}
Let us illustrate this process in the simplest case of $A_1$. The chain of transformations is
\begin{align*}
\epsilon_{\ga\gb} e^{i[p_0\cdot (iz)]}&\rightarrow t_0z^\ga \epsilon_{\ga\gb} e^{i[t_0 p_0\cdot (iz)]}\rightarrow -it_0(iz^\ga-2p_2^\ga) \epsilon_{\ga\gb} e^{i[t_0 (p_0+p_2)\cdot (iz-2p_2)+p_0\cdot p_2]}\rightarrow\\
&\rightarrow -it_0(iz^\ga t_1-2p_2^\ga)z^\gb \epsilon_{\ga\gb} e^{i[t_0 (p_0+p_2)\cdot (t_1 iz-2p_2)+p_0\cdot p_2]}\rightarrow\\
&\rightarrow -t_0(iz^\ga t_1-2p_2^\ga)(iz^\gb-2p_1^\gb) \epsilon_{\ga\gb} e^{i[t_0 (p_0+p_1+p_2)\cdot (t_1 iz-2t_1 p_1-2p_2)+p_0\cdot p_1+p_0\cdot p_2+p_1\cdot p_2]}\rightarrow\\
&\rightarrow 4t_0 (p_1\cdot p_2) e^{i[t_0 (p_0+p_1+p_2)\cdot (-2t_1 p_1-2p_2)+p_0\cdot p_1+p_0\cdot p_2+p_1\cdot p_2]}=\\
&=4t_0 (p_1\cdot p_2) e^{i[p_0\cdot p_1(1-2t_0t_1)+p_0\cdot p_2(1-2t_0)+p_1\cdot p_2(1-2t_0+2 t_0t_1)]}\,,
\end{align*}
where we omit the integral signs and omit the arguments $\omega(y_1)$ and $\omega(y_2)$. Also, in the last but one line we set $z=0$ and in the last line one can easily see the integrand of the Hochschild cocycle. In what follows we will write the equations making this process automatic and more flexible in the choice of a representative.\footnote{It is not hard to build a bi-complex where one differential is the Hochschild one and another one is the de Rham differential in the $z$-space, $dz^\ga \tfrac{\pl}{\pl z^\ga}$. Such resolution will give exactly the FFS cocycle via \eqref{FFSstepbystep}.}

\subsection{Hochschild Cohomology from Shoikhet-Tsygan Formality}
\label{sec:Formality}
Below we just would like to sketch the general relation between higher-spin algebras, Hochschild cocycles and the formality theorems \cite{Kontsevich:1997vb,Tsygan,Shoikhet:2000gw}. The main point is that the proofs of the formality theorems are constructive and provide explicit formulas for the relevant structures.

\begin{wrapfigure}{l}{4.5cm}
\centering\includegraphics[width=.5\linewidth]{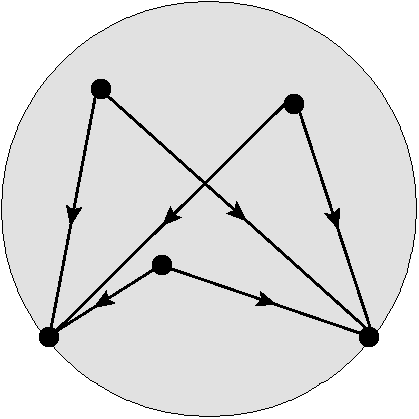}
\caption{Graph for Moyal-Weyl star-product}
\label{figA}
\end{wrapfigure}
It is known, see e.g. \cite{Michel2014}, that  all higher-spin algebras underlying the corresponding gauge theories  can be identified with the $\star$-product algebra of functions on appropriate coadjoint orbits of the $AdS_{d+1}$ group $SO(d,2)$. Since the  coadjoint orbits are symplectic manifolds, the  Fedosov deformation quantization \cite{Fedosov:1994zz} would suffice, in principle, to construct the higher-spin algebra in any dimension. Application of the formality theorems, however,
gives rise to a much richer quantum geometry. Besides the noncommutative algebra of functions it involves the differential forms and polyvector fields forming the full quantum calculus. Loosely,  the role of quantum differential forms is played by the Hochschild  chains, while the graded Lie algebra of polyvector fields is substituted by the differential graded Lie algebra of continuous Hochschild cochains. 
\begin{wrapfigure}{l}{4.5cm}
\centering\includegraphics[width=.5\linewidth]{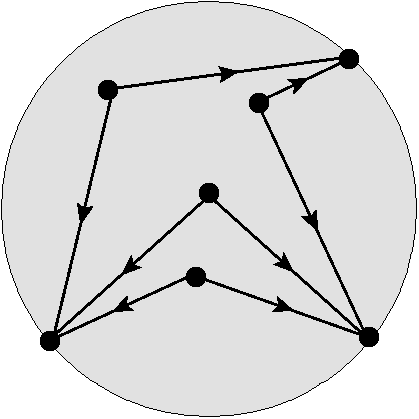}
\caption{$A_1$ graph}
\label{figB}
\end{wrapfigure}The formality map relates the classical and quantum objects in a way respecting all the calculus' operations. 
One can linearize the formality map at any solution to the Maurer-Cartan equation in the Lie algebra of polyvector fields. The corresponding tangent map yields then a homomorphism between the cohomology spaces. Particular solutions to the Maurer-Cartan equation are provided by the Poisson bivectors. Having in mind the higher-spin algebras, one can construct a linearized formality map around the Poisson bivector on a coadjoint orbit of the $AdS$ group.

This must induce an isomorphism between the Hochschild homology of the higher-spin algebra, i.e. the $\star$-product algebra of functions on the coadjoint orbit, and the complex of differential  forms with coboundary operator being given by  
the Lie derivative along the Poisson bivector.\footnote{To some extent this explains the appearance of the auxiliary differential form $\Zf$ 
in the constructions of Section \ref{sec: SE}.} Finally, the dual to the tangent map should relate the corresponding cohomology spaces. 
Effectively, this allows one to write down explicit formulas for nontrivial Hochschild cocycles in terms of the Kontsevich-type integrals, the Shoikhet integrals.

Due to the remarkable isomorphism $SO(3,2)\sim Sp(4)$ the minimal coadjoint orbit of the $AdS$ group is given by the carrier space of the fundamental representation of $Sp(4)$. The residual manifold $\mathbb{R}^4\backslash \{0\}$ appears to be a homogeneous symplectic manifold with respect to the canonical symplectic structure and linear action of $Sp(4)$. The minimal representation corresponds to the $3d$ free conformal scalar field and the higher-spin algebra is the even subalgebra of $A_2$. Algebra $sp(2n)$, which is a subalgebra of $A_n$, does also make sense for applications to higher-spin theories. The interactions are governed by the Hochschild cocycles and the precise specification of the relevant cocycle depends on the details, e.g. what is the realization of the twist map. Nevertheless, whenever the Weyl algebra is at the core of a higher-spin algebra, the relevant cocycle should be the FFS one.
\hfill\phantom{a}
\begin{wrapfigure}{l}{4.5cm}
\centering\includegraphics[width=.5\linewidth]{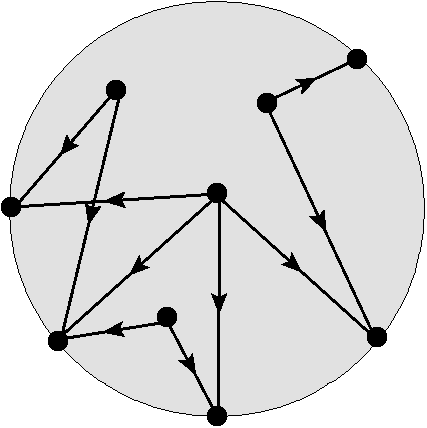}
\caption{$A_2$ graph}
\label{figC}
\end{wrapfigure}
For the minimal coadjoint orbit of $Sp(2n)$ the symplectic structure corresponds to the constant Poisson bivector $C^{\ga\gb}$ and the deformation quantization by means of the formality map gives then the usual Moyal-Weyl $\star$-product. All the configuration space integrals can be done for the Moyal-Weyl case, see Fig.\ref{figA} for the typical graph where each bulk point represents $C^{\ga\gb}$ and the two points on the boundary stay for the arguments in $f\star g$.

The FFS cocycle for the Weyl algebra $A_n$ results from the Shoikhet graphs \cite{Shoikhet:2000gw} and all but the boundary configuration space integrals can be done for the constant Poisson bivector. There still remains a $2n$-fold integral over the ordered points on the boundary, which is what leads to the integral over the simplex \eqref{simplexA}. The typical graphs for $A_1$ and $A_2$ are shown in Fig.\ref{figB},\ref{figC}, where the boundary points correspond to the arguments of the cocycle and the central vertex to the $\epsilon$-symbol.

One may expect that the same technique of generating nontrivial Hochschild cocycles for higher-spin algebras works for other cases as well, even though the corresponding  symplectic structures and the group actions may not be linear anymore.\footnote{See e.g. \cite{Gunaydin:2016bqx} for the review of the quasi-conformal approach that allows one to realize the minimal unitary representations by a minimal number of oscillators.}

\section{Higher Higher-Spin Theories}
\label{sec:HigherHS}
Inspired by the existence of the Hochschild cocycles of the Weyl algebra $A_{n}$ for any $n$, we would like to find unfolded equations where such cocycles could serve as interaction vertices. We also construct Vasiliev-like equations that generate such cocycles upon solving for the extra variables. As it was already mentioned in the Introduction, $sp(2n)$-symmetries are often occur in the higher-spin context and we expect the systems below to be of some interest not only as a tool for generating nontrivial Hochschild cocycles. The $4d$ Vasiliev equations result from a special truncation that is possible for $A_1$. In particular, we show how to dissect the Vasiliev equations in such a way that only the $\mathcal{V}(\omega,\omega,C)$-vertex is generated and no more nonlinearities in $C$ are needed. This also gives description of higher-spin fluctuations over a higher-spin background. The $sp(4)$ system may be relevant for constructing the on-shell action for the $4d$ HS theory, see also \cite{Vasiliev:2015mka}.

\subsection{Unfolded Realization of the Hochschild Cocycles}
\label{sec:UnfldHochReal}
As a starting point we assume the existence of a nontrivial function $\Phi(a_1,...,a_{2n})$ of $2n$ variables that obeys the Hochschild cocycle condition for some HS algebra:
\begin{align}\label{spgenHochschild}
   - a_1\star \Phi(a_2,...,a_{2n},a_{2n+1})+\Phi(a_1\star a_2,...,a_{2n},a_{2n+1})-... +\Phi(a_1,a_2,...,a_{2n})\star \tilde{a}_{2n+1}=0\,.
\end{align}
Here we do not have yet to assume that the HS algebra is the Weyl algebra $A_{n}$, though we will make such an assumption in the next Section. First of all, the simplest unfolded equations where $\Phi$ can be used as an interaction vertex\footnote{In this Section, `interaction vertex' simply means something nonlinear in fields.} require a $(2n-1)$-form $U$ in the twisted-adjoint representation of the HS algebra:
\besubeqs\label{systemA}
\begin{align}
dU&=\omega \star U +U\star \tilde \omega+\Phi(\omega,...,\omega)\,,\\
d\omega&=\omega\star \omega\,.
\end{align}
\esubeqs
The Frobenius integrability of this equation, bearing in mind the passage from the Lie to associative structures, is exactly the Hochschild cocycle condition. A natural extension of the system \eqref{systemA} is to add zero-forms $K$ that are in the adjoint representation of the HS algebra:
\besubeqs\label{systemB}
\begin{align}
dU&=\omega \star U +U\star \tilde \omega+\mathcal{V}(\omega,...,\omega,K)\,,\\
d\omega&=\omega\star \omega\,,\\
dK&=\omega \star K-K\star \omega\,.
\end{align}
\esubeqs
The equivariance condition, c.f. \eqref{Equivariance}, implies that 
\begin{align}
\mathcal{V}(\omega,...,\omega,K)&=\Phi(\omega,...,\omega)\star K\,.
\end{align}
Fields in the complementary representations, i.e. zero-forms in the adjoint and one-forms (or higher forms) in the twisted-adjoint representations naturally occur in the higher-spin context. When $\omega$ is an $AdS$-connection, the equation for $K$ decomposes into an infinite set of equations for Killing tensors. Whenever some unfolded equations are available one can realize the $\pi$-automorphism as an inner one by enlarging the set of generators. The equations remain consistent but the field content is at least doubled with additional fields taking values in the complementary representations, see e.g. \cite{Vasiliev:1999ba}. In practice, one usually tries to get rid of such fields as they can mix with the physical ones, see e.g. \cite{Kessel:2015kna}.

A more natural, in the HS sense, way to take advantage of $\Phi$ is to introduce a $(2n-1)$-form $G$ in the adjoint representation of the HS algebra, while the discrepancy in the type of representations can be compensated by introducing a zero-form $C$ in the twisted-adjoint representation:
\besubeqs\label{systemC}
\begin{align}
dG&=\omega \star G +G\star  \omega+\mathcal{V}(\omega,...,\omega,C)\,,\\
d\omega&=\omega\star \omega\,,\\
dC&=\omega \star C-C\star \tilde\omega\,.
\end{align}
\esubeqs
The equivariance equation, c.f. \eqref{Equivariance}, implies that\footnote{Cocycles with higher powers of $C$ may exist in the formal sense and are obtained by adding more powers of $C\star \tilde{C}$. However, it is worth stressing that those are non-local when the formal unfolded equations are turned into differential equations. See footnote \ref{higherorders}.} 
\begin{align}
\mathcal{V}(\omega,...,\omega,C)&=\Phi(\omega,...,\omega)\star \tilde C\,.
\end{align}
From the Hochschild cocycle vantage point the need for $C$ is to merely adjust the type of representation.

All the unfolded equations above are complete in the sense that no nonlinear corrections are needed, which is due to the fact that there is no backreaction: the fields $G$ or $U$ that the Hochschild cocycle makes a source to are different from the arguments of the cocycle. This can also be applied to the case $n=1$.

\paragraph{Higher-Spins on Background of Their Own.} It turns out that the case of the Hochschild cocycle being a two-cocycle is special. In this case $G$ is a one-form and, after renaming $G$ to $\omega$ and $\omega$ to $\Omega$, the system reads 
\besubeqs\label{linfluct}
\begin{align}
d\Omega&=\Omega\star \Omega\,,\\
d\omega&=\Omega \star \omega +\omega\star  \omega+\mathcal{V}(\Omega,\Omega,C)\,,\\
dC&=\Omega \star C-C\star \tilde\Omega\,,
\end{align}
\esubeqs
where $\mathcal{V}(\Omega,\Omega,C)=\Phi(\Omega,\Omega)\star \tilde{C}$. Therefore, $\omega$ is of the same nature as $\Omega$. It is important to stress that  no higher-order terms are needed. In the context of the $4d$ HS theory, see Section \ref{sec:freedata}, whenever $\Omega$ is a non-degenerate flat connection of $sp(4)$, it defines $AdS_4$. It is clear that $\omega$ and $C$ form the linearized unfolded equations reviewed in Section \ref{sec:freedata}, which describe free higher-spin fields on $AdS_4$. The equations make perfect sense for any flat $\Omega$. As a result, we have exact equations that describe the propagation of higher-spin fields over a higher-spin background. The equations are local and thereby make sense as differential equations and not only as formal ones. Due to the importance of such equations we write them down explicitly in Appendix \ref{app:fluctuations}. Clearly, the equations above are not bound to $4d$ and work the same way for any higher-spin theory in any other dimension $d\geq4$ whenever the Hochschild cocycle can be written down or proved to exist. Also, we can write down the global higher-spin algebra transformations:
\besubeqs
\begin{align}\label{fronsdaltrans}
\delta \omega&= \xi\star \omega-\omega\star \xi +\mathcal{V}(\xi,\Omega,C)-\mathcal{V}(\Omega,\xi,C)\,,\\
\delta C&= \xi\star C+C\star \tilde{\xi}\,,\label{fronsdaltransB}
\end{align}
\esubeqs
where $\xi$ are the `Killing tensors', i.e. the symmetries of the vacuum that obey 
\begin{align}
\delta \Omega\equiv d\xi-\Omega\star \xi+\xi\star \Omega=0\,.
\end{align}
Eq.\eqref{fronsdaltransB} shows that the zero-forms belong to the twisted-adjoint representation of the HS algebra. The most interesting part is in Eq.\eqref{fronsdaltrans}: roughly speaking, the Fronsdal fields belong to the adjoint of the HS algebra with the correction due to the Hochschild cocycle hidden in $\mathcal{V}$. An advantage of the unfolded approach is that this nontrivial transformation law is obtained just by replacing one $\Omega$ with $\xi$ in \eqref{linfluct}, in accordance with the general rule \eqref{unfldgaugesymmetry}. 

As a final remark, it may be interesting to make the fields backreact onto themselves, which should call for the non-linear completion of the system. An infinite series of terms can only come from expansion in zero-forms. Having enough zero-forms one can use the Hochschild cocycle as a source for the forms with degrees all the way down to zero:
\besubeqs
\begin{align}
dG_{2n-1}&=\omega \star G_{2n-1} +G_{2n-1}\star  \omega+\mathcal{V}(\omega,...,\omega,C)\,,\\
dG^{\aAs}_{2n-2}&=\omega \star G^{\aAs}_{2n-2} -G^{\aAs}_{2n-2}\star  \omega+C^\aAs \frac{\delta}{\delta \omega}\mathcal{V}(\omega,...,\omega,C)\,,\\
....\,,\\
d\omega&=\omega\star \omega+...\,,\\
dC^\aAs&=\omega \star C^\aAs-C^\aAs\star \tilde\omega+...\,.
\end{align}
\esubeqs
The idea of this system is to replace step-by-step one-forms $\omega$ with zero-forms $C^\aAs$, which allows us to use the same Hochschild cocycle as a vertex for forms of lower degree. When we reach one-forms $G_1$ we can identify those with $\omega$, which makes fields backreact.

\subsection{Vasiliev Resolution}
\label{sec:VasDouble}
On one hand we have a number of interesting unfolded equations where the Hochschild cocycles constitute the interaction terms. On the other hand it was observed in Section \ref{sec:Delta} that the Hochschild cocycle can be thought of as being a coboundary in a space with the doubled set of generators $y_\ga,z_\ga$ and with a nontrivial star-product that mixes $y$ and $z$; we called this the Vasiliev double. Also, the Vasiliev equations can be thought of as a device to build the Hochschild cocycle for $A_1$ and a non-linear completion of it. 

In this Section we construct the equations that: (i) are simple enough; (ii) are easy to check to be consistent; (iii) over an appropriate vacuum they give first-order differential equations with respect to $z_\ga$; (iv) upon solving these equations one generates the Hochschild cocycle for any $A_{n}$. The cocycle condition results from the general consistency of the system. The non-triviality follows from the fact that the interactions cannot be redefined away. Mathematically, we construct the injective resolution of the Hochschild complex in the sense of Cartan-Eilenberg \cite{CartEil}. This allows one to arrive at the Hochschild cocycle in much simpler way.

We begin with rigorous definitions of the algebra and then discuss the simplest instance of the equations that do not have zero-forms \eqref{systemA} and then turn to more interesting cases \eqref{systemB}, \eqref{systemC}.

\subsubsection{Algebraic Preliminaries}
\label{sec:}
Firstly, we define the algebra of differential forms in $x$-$z$ space that are also functions of $x,y,z$ and discuss various natural operations.\footnote{Similar objects --- extensions of the Vasiliev equations with higher forms in $x$ and $z$ spaces --- have already appeared in recent papers \cite{Boulanger:2011dd,Boulanger:2015kfa,Vasiliev:2015mka,Didenko:2015cwv}.} Let us introduce the bigraded associative algebra of differentials forms $A=\bigoplus A^{p,q}$, whose generic element reads
\begin{equation}\label{ge}
a=  a(y,z|x|dx,dz)=a_{\mm_1\cdots \mm_p\alpha_1\cdots \alpha_q}(y,z|x)dx^{\mm_1}\wedge\cdots\wedge dx^{\mm_p}\wedge dz^{\alpha_1}\wedge \cdots\wedge dz^{\alpha_q}\in A^{p,q}\,.
\end{equation}
Here the coefficients $a_{\mm_1\cdots \mm_p\alpha_1\cdots\alpha_q}$ are assumed to be smooth functions in $x$'s and polynomial in $y$'s and $z$'s. The associative product in $A$, denoted by $\star$, combines the usual exterior product of forms and the star-product of $z$'s and $y$'s:\footnote{As it was already mentioned in Section \ref{sec:Double}, the star-product that allows one to generated the Hochschild cocycle is not unique and we choose the simplest representative introduced in \cite{Vasiliev:1992av}. See also \cite{Alkalaev:2014nsa} for some variations.  }
\begin{equation}\label{yzstar}
(a\star b)(y,z) =a(y,z) \exp i\left( \frac{\overleftarrow{\partial}}{\partial y^\alpha} +\frac{\overleftarrow{\partial}}{\partial z^\alpha}\right)C^{\alpha\beta}\left( \frac{\overrightarrow{\partial}}{\partial y^\beta}-\frac{\overrightarrow{\partial}}{\partial z^\beta}\right)b(y,z)\,.
\end{equation}
The star-product is the Moyal-Weyl one for $y$ and $z$ separately and is normal ordered with respect to $y\pm z$. As the whole construction depends crucially on doubling of the variables from $y$ to $y,z$ and on some of the properties of this star-product we refer to it as the Vasiliev double.

The algebra $A$ is unital and $1\star a=a\star 1$ for all $a\in A$ and $1\in \mathbb{C}$. Associated to the bigrading is the total grading  $A^m=\bigoplus_{p+q=m} A^{p,q}$. The algebra $A$ admits the pair of anti-commuting differentials: $d$ and $\delta$, such that $d\delta+\delta d=0$ and
\begin{align}\label{dd}
\begin{aligned}
d&: A^{p,q}\rightarrow A^{p+1,q}\,, \qquad\qquad&& d=dx^i\wedge \frac{\partial}{\partial x^i}\,,\qquad\qquad&&d^2=0\,,\\ \delta&:A^{p,q}\rightarrow A^{p,q+1}\,,\qquad\qquad && \delta=dz^\alpha\wedge \frac{\partial}{\partial z^\alpha}\,,\qquad\qquad&&\delta^2=0\,.
\end{aligned}
\end{align}
The differentials make $A$ into a bicomplex. It is clear that 
\begin{align}
d(a\star b)&=da\star b+(-1)^{m}a\star db\,,&& \delta(a\star b)=\delta a\star b+(-1)^{m}a\star \delta b\,,
\end{align}
for all $a\in A^{m}$ and $ b\in A$. One of the key ingredients that will explain some, otherwise strange, sign factors in the equations is to observe that the bigraded, bidifferential algebra $(A,d,\delta)$ admits an involutive automorphism  $\pi: A\rightarrow A$ defined as
\begin{equation}\label{bigtwist}
(\pi a)(x,y,z|dx,dz)=a(x,-y,-z|dx,-dz)\,.
\end{equation}
It is easy to see that for all $a,b\in A$ we have
\begin{align}
\pi(a\star b)&=\pi(a)\star \pi(b)\,, & \pi^2&=id\,, &
\pi(d a)&=d\pi(a)\,,& \pi(\delta a)&=\delta \pi(a)\,.
\end{align} 
Using the $\star$-product above and the involution $\pi$, we can define the usual commutator  and the $\pi$-commutator as follows: 
\begin{align}
&\begin{aligned} [a,b]_{\phantom{\pi}}&=a\star b-(-1)^{nm}b\star a\,,\\
[a,b]_\pi&=a\star b-(-1)^{nm}b\star \pi(a)\,, 
\end{aligned} &&\forall a\in A^n\,,\qquad \forall b\in A^m\,.
\end{align}
The commutator makes $A$ into the Lie superalgebra $L(A)$ endowed with the adjoint action $ad: L(A)\rightarrow End (A)$. The $\pi$-commutator gives rise to one more representation $ad^\pi: L(A)\rightarrow End(A)$, called twisted-adjoint. By definition, 
\begin{align}
L(A)\ni a \mapsto ad^\pi_a&:A\rightarrow A\,,&&  ad^\pi_a b=[a,b]_\pi\,,&& \forall b\in A\,.
\end{align}
Unlike the adjoint the twisted-adjoint action is not a derivation of the associative algebra $A$. 

In practice, we will have to deal with certain non-polynomial elements of the star-product algebra and it has always been appreciated that a product of two such elements can be ill-defined. In the Vasiliev equations, which include the backreaction of fields onto themselves, such potentially dangerous products do appear and a proof was given that no infinities arise in the formal perturbation theory, see e.g. \cite{Vasiliev:1990vu,Vasiliev:1999ba}. In the case under consideration a weaker assumption will suffice.

Denote by $\hat {A}$ the completion of the space $A$. This is given by the differential forms \eqref{ge} with coefficients being smooth functions in $x$'s and {\it {formal power series}} in $y$'s and $z$'s. Note that unlike $A$ the completion $\hat{A}$ is not an algebra with respect to $\star$-product \eqref{yzstar} due to the possible divergences. Nonetheless, $\hat{A}$ can still be viewed as a bimodule over $A$.

\subsubsection{Simplest Equations}
\label{sec: SE}
Let us take two elements $\Wf=\Wf_\mm(y,z|x) dx^\mm\in A^{1,0}$ and $\Sf=\Sf_\ga(y,z|x)dz^\ga\in A^{0,1}$, which will serve as connections along $x$ and $z$. The boldface letters will be used to denote fields taking values in the full algebra of $(y,z)$, i.e. in the Vasiliev double, while the usual letters are reserved for the $z$-independent fields like $\omega$, $C$. We can endow the bigraded space $\hat{A}$ with the operators 
\besubeqs
\begin{align}
\Df_\Wf\,,\Dfa_\Wf&: \hat{A}^{p,q}\rightarrow \hat{A}^{p+1,q}\,,&  \Df_\Wf&=d-ad^\pi_\Wf\,,& \Dfa_\Wf&=d-ad_\Wf\,,\\[3mm]
\Df_\Sf\,,\Dfa_\Sf&: \hat{A}^{p,q}\rightarrow \hat{A}^{p,q+1}\,, & \Df_\Sf&=\delta-ad^\pi_\Sf\,,& \Dfa_\Sf&=\delta-ad_\Sf\,,
\end{align}
\esubeqs
as well as their sums
\begin{align}
\Df\,,\Dfa&: \hat{A}^m\rightarrow \hat{A}^{m+1}\,,& \Df&=\Df_\Wf+\Df_\Sf\,,& \Dfa&=\Dfa_\Wf+\Dfa_\Sf\,.
\end{align}
Here the action of the operators $d$, $\delta$, $ad_a$ and $ad^\pi_a$ naturally extends from $A$ to the bigger space $\hat{A}$. 

Using the geometric language, we will refer to $\Df$ as the connection in $\hat A$. Then the curvature $\Rf\in A^2$ of the connection $\Df$ 
is defined  in the usual way, 
$\Df^2=-ad^\pi_\Rf$, or $\Dfa^2=-ad_\Rf$ for $\Dfa$, which gives the same curvature, of course. It has three different components: along $dx\wedge dx$, $dx\wedge dz$ and $dz\wedge dz$. 

Following the Fedosov terminology \cite{Fedosov:1994zz}, we say that $\Df$ or $\Dfa$ is an {\it abelian connection} if $\Rf\in \ker ad^\pi$ or $\Rf\in \ker ad$ and we call it flat if $\Rf=0$. The zero-curvature condition $\Rf=0$ amounts to 
\begin{align}
    d\Wf&=\frac12[\Wf,\Wf]\,,& \delta \Wf + d\Sf-[\Sf,\Wf]&=0\,,& \delta \Sf&=\frac12[\Sf,\Sf]\,.
\end{align}
Any abelian connection $\Df$ makes $\hat A$ into a cochain complex with respect to the total degree. 

In the case of the usual adjoint action the center of the Weyl algebra is known to be constants. On contrary, the center of the twisted-adjoint action is nontrivial and the kernel of the operator $\Df: A\rightarrow A$ is obviously nonzero. In particular, the center contains the element  
\begin{align}\label{zcycle}
\Zf&=\varkappa\, dz^{1}\wedge\cdots\wedge dz^{{2n}}\in A^{0,2n}\,,&&\varkappa=e^{iy^\nu z_\nu}\,, & \Df\Zf&=0\,,
\end{align}
where $\varkappa$ is known as the inner Klein operator \cite{Vasiliev:1992av}.\footnote{If we keep track of the Planck constant $\hbar$, which enters the star-product as $\exp i[\hbar\overleftarrow{\pl}\cdot \overrightarrow{\pl}]$, we find that the Klein operator is of quasi-classical nature $\exp i\left[\tfrac{1}{\hbar} z\cdot y\right]$.} The Klein operator satisfies
\begin{align} \label{kleinchik}
\varkappa\star \varkappa&=1\,, &&\varkappa\star f(y,z)\star \varkappa=f(-y,-z)\,, && f(y,z)\star \varkappa=f(-z,-y)\varkappa\,,
\end{align}
and thereby realizes the twist map \eqref{bigtwist} on $y,z$ as an inner automorphism. Note that it does not act on $dz$. One may ask whether the $2n$-cocycle $\Zf$ is nontrivial. To answer this question let us consider the equation
\begin{align}\label{justcocycle}
\Df\Uf=\Zf\,,
\end{align}
where the l.h.s. makes a set of forms $\Uf$ to transform in the twisted-adjoint representation. Expanding $\Uf$ in homogeneous components,
\begin{align}
\Uf=\Uf_0+\Uf_1+\cdots+\Uf_{2n-1}\,,
\end{align}
where $\Uf_k\in A^{2n-1-k,k}$, we get the following chain of equations:
\besubeqs
\begin{align}
\Df_\Sf \Uf_{2n-1}&=\Zf\,, \\[3mm]
\Df_\Sf \Uf_{k-1}&=-\Df_\Wf\Uf_k\,, \qquad\qquad k=1,\ldots, 2n-2\,, \\[3mm] 
\Df_\Wf\Uf_0&=0\,.
\end{align}
\esubeqs
Connection $\Sf$ being flat, let us write the equations in the gauge $\Sf=0$. Firstly, we find that
\begin{align}
\delta \Wf&=0\,,&& d\Wf=\frac12[\Wf,\Wf]\,,
\end{align}
i.e. $\Wf$ is $z$-independent and we can recover the flat connection $\omega$ via $\Wf=\omega(y|x)$. For the rest of the equations we find
\besubeqs\label{dG}
\begin{align}
\delta \Uf_{2n-1}&=\Zf\,,\\[3mm]
\delta \Uf_{k-1}&=-\Df_\omega \Uf_k\,,\qquad\qquad k=1,\ldots, 2n-2\,, \label{zzero}\\[3mm]
\Df_\omega\Uf_0&=0\,. \label{dGlast}
\end{align}
\esubeqs
It is worth emphasizing that $\Df_\omega$ acts nontrivially on functions of $z$, c.f. \eqref{newstar}, and to large extent this is the most important part of this action. In order to solve the equations we introduce the standard contracting homotopy operator $\delta^{-1}$ such that $\delta\delta^{-1}+\delta^{-1}\delta=1$. In practice, having a $\delta$-closed $q$-form in the $z$-space $\Ff\equiv \Ff_{\ga_1...\ga_q}(z) dz^{\ga_1}\wedge...\wedge dz^{\ga_q}$ we can define $\delta^{-1}$ as  
\begin{align}
(\delta^{-1} \Ff)_{\ga_2...\ga_q}(z)\,dz^{\ga_2}\wedge...\wedge dz^{\ga_q} &= z^\nu\int_0^1dt\, t^{q-1}\Ff_{\nu\ga_2...\ga_q}(zt)\,dz^{\ga_2}\wedge...\wedge dz^{\ga_q}\,.
\end{align}
Here the possible dependence of $y$, $x$ and $dx$ is irrelevant. Now we can solve all but one equations in the following way:
\besubeqs
\begin{align}
\Uf_{2n-1}&=\delta^{-1} \Zf +\delta \Qf_{2n-2}\,, \\[3mm]
\Uf_{2n-m-1}&=(-\delta^{-1} \Df_\omega)^{m}(\delta^{-1} \Zf) +\Df_\omega \Qf_{2n-m-1}+\delta \Qf_{2n-m-2}\,, && m=1,\ldots, 2n-2\,,\\[3mm]
\Uf_0&=(-\delta^{-1} \Df_\omega)^{2n-1}(\delta^{-1} \Zf) +\Df_\omega \Qf_{0} +U\,,
\end{align}
\esubeqs
where $\Qf_k\in A^{2n-k-2,k}$ parametrize the solution of the homogeneous equations, or, in other words, $\delta$-exact forms. In the solution to the last equation there is a $z$-independent function $U$, which by the form-degree counting argument must be a $(2n-1)$-form in the $x$-space. $U$ represents $\delta$-cohomology, which is concentrated in degree-zero forms in the $z$-space. Using the  relations 
\begin{align}
d\delta^{-1}+\delta^{-1} d&=0\,,&& \delta \Df_\omega + \Df_\omega\delta =0\,,&& (\delta^{-1})^2=0\,,
\end{align}
we observe that the $d$-part of $\Df_\omega$ does not contribute at all, i.e. the solution for $\Uf_0$ is 
\begin{align}
\Uf_0&= \Pf+\Df_\omega \Qf_{0} +U\,, && \Pf\equiv(\delta^{-1} ad^\pi_\omega)^{2n-1}(\delta^{-1} \Zf)\,.
\end{align}
Finally, we need to substitute this into the last equation \eqref{dGlast}. We note that $\Df_\omega \Uf_0$ does not depend on $z$. Indeed, we can see that 
\begin{align}
\delta \Df_\omega \Uf_0=-\Df_\omega \delta \Uf_0=\Df_\omega  \Df_\omega \Uf_1\equiv0
\end{align}
thanks to the flatness of $\omega$. The $\Df_\omega \Qf_{0}$ part of $\Uf_0$ represents the standard gauge transformations and does not contribute to $\Df_\omega \Uf_0$ due to $\Df_\omega^2=0$. The $U$-part is already $z$-independent. Therefore, the only nontrivial statement is that $\Df_\omega \Pf$ is $z$-independent. In any case, we can set $z=0$ in the last equation since it has been proven to be $z$-independent:
\begin{align}\label{DW}
\Df_\omega U=-\Df_\omega\Pf \Big|_{z=0}=ad^\pi_\omega \Pf \Big|_{z=0}\,,
\end{align}
where we used that $d\delta^{-1}\ldots |_{z=0}=0$. As a result, the equation reads
\begin{align}\label{FFSunfld}
dU&= [\omega, U]_\pi +\Phi(\omega,\ldots,\omega)\,, && \Phi(\omega,\ldots,\omega)=ad^\pi_\omega(\delta^{-1} ad^\pi_\omega)^{2n-1}(\delta^{-1} \Zf)\Big|_{z=0}\,.
\end{align}
Together with the flatness of $\omega$ we discover the equations of type \eqref{systemA}. Applying the operator $\Df_\omega$ to both sides of (\ref{DW}) yields the identity
\begin{equation}
[\omega, \Phi(\omega,\ldots,\omega)]_\pi=\frac12\sum_{i=1}^{i=2n}  (-)^{i+1}\Phi(\omega,\ldots,\underbrace{[\omega,\omega]}_i,\ldots,\omega)\,,
\end{equation}
where we used that $\omega$ is flat. In other words, $\Phi$ is a $2n$-cocycle of the Lie algebra $L(A_n)$ with coefficients in the twisted-adjoint representation. Taking advantage of the fact that the equations make sense for matrix-valued fields as well, we can always go from $L(A_n)$ to $L(A_n\otimes \mathrm{Mat})$ and with the help of Appendix \ref{app:Hochschild} this shows that the cocycle above is equivalent to the FFS Hochschild cocycle of Section \ref{sec:FFS}. In order to get exactly the FFS cocycle (its anti-symmetric part) one should adjust the gauge ambiguity represented by the $\delta Q_k$-terms. It can be shown that the representative \eqref{FFSunfld} is not $sp(2n)$-basic, c.f. \eqref{spbasic}. 

We would like to evaluate the cocycle on the simplest connection that is linear in $y$: $\omega=\xi^\ga_\mm y_\ga dx^\mm$. It is {\it abelian} whenever $\xi^\ga_\mm$ is $x$-independent, i.e. its curvature belongs to the center. One can make it into the flat connection
\begin{align}\label{linearcon}
\omega&=\xi^\ga_\mm y_\ga dx^\mm+i \xi^\ga_\mm \xi^\gb_\nn x^\mm dx^\nn C_{\ga\gb}\,.
\end{align}
The cocycle gives
\begin{align}
\Phi(\omega,...,\omega)&= \frac{1}{2n!} \xi^{\ga_1}\wedge...\wedge\xi^{\ga_{2n}} \epsilon_{\ga_1\ldots \ga_{2n}}\,.
\end{align}
Let us explain why the cocycle $\Phi$ is nontrivial. The first evidence is that the  cocycle, which is $z$-independent by construction, is obtained as $\Df_\omega$ of a potential $\Pf$ that is $z$-dependent, \eqref{DW}. This suggests that it cannot be represented as $\Df_\omega$ of anything that is $z$-independent. Indeed, if it were the case we would find from \eqref{zzero}
\begin{align}
ad^\pi_\omega(\delta^{-1} ad^\pi_\omega)^{2n-2}(\delta^{-1} \Zf)\Big|_{z=0}=0\,.
\end{align}
This is disproved by evaluating it on \eqref{linearcon}, which gives a multiple of  $\xi^{\ga_1}\wedge...\wedge\xi^{\ga_{2n-1}} \epsilon_{\ga_1\ldots \ga_{2n-1}\nu}dz^\nu$. Another side of the same coin is that Eqs.\eqref{dG} are nontrivial because the initial system has an `interaction term' $\Zf$ that cannot be redefined away. Had we chosen $\Zf$ to be form of degree less than $2n$ we would have found a nontrivial constraint $\delta \Zf=0$. Thanks to $\Zf$ being a top-form in the $z$-space we have $\delta \Zf\equiv0$.

\subsubsection{Equations with Zero-Forms}
\label{sec:doublezero}
We now turn to the systems which are reminiscent of the unfolded equations for higher-spin fields in that there are zero-forms. The first system to reproduce is \eqref{systemB}. The proposal is
\besubeqs
\begin{align}
\Df^2&=0\,, \\[3mm]
\Df \Uf &= \Zf\,, & \Zf&= (\Kf\star \varkappa)\, dz^{1}\wedge\cdots\wedge dz^{{2n}}\,,\\[3mm]
\Dfa \Kf&=0\,,
\end{align}
\esubeqs
where $\Kf\in A^{0,0}$, i.e. it is a zero-form. $\Zf$ is $\Df$-closed due to $\Dfa \Kf=0$. The consistency requires only that $\Dfa_\Wf \Kf=0$ because $\Zf$ is a top-form in the $z$-space and its closure is trivial. Here we can use 
\begin{align}
\Df_\Wf (\Kf\star \varkappa)=(\Dfa_\Wf \Kf)\star \varkappa\,, \end{align}
i.e. the twisted-adjoint derivative is mapped to the adjoint one. However, it is important to impose $\Dfa \Kf=0$, which is consistent. Upon choosing the simplest $\Sf=0$ gauge we again find $\Wf=\omega$ and, in addition,
\begin{align}
dK&=[\omega, K]\,, && \Kf=K(y|x)\,,&& \delta \Kf=0\,.
\end{align}
The rest of the analysis does not change and at the end of the $\Uf$-chain we find
\begin{align}\label{FFSunfldB}
dU&= [\omega, U]_\pi +\mathcal{V}(\omega,\ldots,\omega,K)\,, && \mathcal{V}(\omega,\ldots,\omega,K)=ad^\pi_\omega(\delta^{-1} ad^\pi_\omega)^{2n-1}(\delta^{-1} \Zf)\Big|_{z=0}\,,
\end{align}
which is a system of type \eqref{systemB}.

We now turn to the system \eqref{systemC} that contains zero-form $C$ in the twisted-adjoint representation of the higher-spin algebra. The proposal is
\besubeqs\label{vasC}
\begin{align}
\Dfa^2&=0\,, \\[3mm]
\Dfa \Gf &= \Zf\,, & \Zf&= (\Bf\star \varkappa)\, dz^{1}\wedge\cdots\wedge dz^{{2n}}\,,\\[3mm]
\Df \Bf&=0\,.
\end{align}
\esubeqs
Here $\Gf\in A^{2n-1}$ and $\Bf\in A^0$. The trick is the opposite: the $\Dfa$-closure of $\Zf$ does not lead to any constraints on $\Bf$ along the $z$-direction but imposes $\Df_\Wf \Bf=0$. The latter equation can be supplemented with $\Df_\Sf \Bf=0$. Again in the $\Sf=0$ gauge we can associate $\Wf$ with $\omega$ and find 
\begin{align}
dC&=[\omega, C]_\pi\,, && \Bf=C(y|x)\,,&& \delta \Bf=0\,.
\end{align}
After lifting $\Zf$ to the last component $\Gf_0$ of $\Gf$, which belongs to $A^{2n-1,0}$, we have
\begin{align}\label{FFSunfldC}
dG&= [\omega, G] +\mathcal{V}(\omega,\ldots,\omega,C)\,, && \mathcal{V}(\omega,\ldots,\omega,C)=ad^\pi_\omega(\delta^{-1} ad^\pi_\omega)^{2n-1}(\delta^{-1} \Zf)\Big|_{z=0}\,,
\end{align}
where $\delta G=0$ represents the $\delta$-cohomology that contributes to the solution for $\Gf_0$. In such a way we recover \eqref{systemC}.

\subsubsection{Comments}
\label{sec:comments}
Let us briefly discuss some general features of the equations introduced above. 

\paragraph{Matrix Extensions.} Firstly, all of the systems can be generalized by letting fields take values in the Lie algebra built out of the tensor product of the Vasiliev double and matrix algebra. This fact, see also Section \ref{sec:UnfldHoch} and Appendix \ref{app:Hochschild}, allows us to claim that the cocycles obtained above are equivalent to a somewhat simpler FFS cocycle of Section \ref{sec:FFS}.

\paragraph{Absorbing $\boldsymbol{\delta}$.} The equations we discussed suffice to generate nontrivial Hochschild cocycles. This does not require, at least in the gauge that we employed, to use any properties of the star-product in the $z$-space. Nevertheless, if we postulate the star-product to be the one in \eqref{yzstar}, then $\delta$ can be absorbed into $\Sf$ as a vacuum value. Indeed, $[z_\ga,f]=-2i\pl^z_\ga f$.

\paragraph{Higher-orders.} We reproduced the unfolded equations \eqref{systemA}, \eqref{systemB}, \eqref{systemC} with the Hochschild cocycle as a vertex. However, the perturbative analysis does not stop here despite the fact that we do not expect any corrections. It is easy to see that this is indeed so. First of all, we note that $\Sf_\ga\,dz^\ga$ has no corrections at the first order. $\Wf$ does have corrections, but they are given by a $z$-independent function $\omega_1$. To cut long story shot, the perturbation theory, say for the last system, leads to
\begin{align}
\Wf&= \omega(y)+\omega_1(y)+...\,, &
\Bf&=0+C_1(y)+C_2(y)+...\,, &
\Sf_\ga&= 0\,,
\end{align}
i.e. to over-parametrization of $\omega(y)$ and $C(y)$. Therefore, we can truncate at the first order or sum up the fluctuations into full $\omega(y)$ and $C(y)$, for which we find \eqref{systemC}.

\paragraph{Relation to the Vasiliev Equations.} 
As compared to Eqs.\eqref{systemC} of Section \ref{sec:UnfldHochReal}, there is an important degeneracy in the case of $A_1$: there are only two components in the $\Gf$-field, $\Gf_\ga dz^\ga$ and $\Gf_\mm dx^\mm$ and they look similar to $\Sf_\ga dz^\ga$ and $\Wf_\mm dx^\mm$. In fact, they can be identified, which reduces the field content and also induces additional nonlinearities in the perturbation theory. This way we recover the holomorphic truncation of the $4d$ Vasiliev equations reviewed in Appendix \ref{app:Vasiliev}:
\begin{align} \label{holovas}
\Dfa^2&=\frac12 (\Bf\star \varkappa)\,dz_\nu \wedge dz^\nu\,, &
\Df \Bf &= 0\,.
\end{align}
In this case the $\Zf$-term moves to the r.h.s. of the zero-curvature equation, i.e. $\Rf=\Zf$ and $\Dfa \Zf=0$, which does not spoil the consistency. The full $4d$ equations are obtained by simply attaching the second copy of the $(y_\ga,z_\ga)$ variables, $(\bry_\gad,\brz_\gad)$, and extending $\Sf_\ga$ with $\Sf_\gad$, which also requires one more Klein operator $\brklein=e^{i\bry^\gnd \brz_\gnd}$ and certain kinematical constraints, see Appendix \ref{app:Vasiliev}. The equations of this Section can also be generalized by extending $y,z$ to several families $y^i,z^i$ and $\Sf$ to $\Sf^i$. 

The interpretation of the trick above is as follows. Eqs.\eqref{systemC} originate from system \eqref{vasC} and describe linear fluctuations of higher-spin fields over a HS background given by any flat connection. If the background is just $AdS$ then they are equivalent to the free Fronsdal equations and contain no information about possible interactions. Here we have the equations that are valid over much more general backgrounds that probe all HS symmetries and it seems that this knowledge is enough to reconstruct the theory in one goal by identifying background and fluctuating fields in the generating equations \eqref{vasC}.\footnote{We are grateful to Xavier Bekaert and Maxim Grigoriev for this interpretation.}

\paragraph{$\boldsymbol{A_2}$-cocycle.} The case of $A_2$ is also of interest. The $4d$ bosonic higher-spin algebra is the even subalgebra of $A_2$. Therefore, the $A_2$ FFS cocycle gives an interesting four-form that is constructed out of $\omega$, (or $\omega$ and $C$ if we consider \eqref{systemC}). This may be interpreted as a part of an on-shell Lagrangian, which can be used for establishing the AdS/CFT  correspondence. At free level such a Lagrangian was recently proposed in \cite{Sharapov:2016qne}. 
It should also arise from the extended system of \cite{Vasiliev:2015mka}. We evaluate the $A_2$-cocycle in Appendix \ref{app:spfour}.

\paragraph{Remark on the Formality.} It should note escape one's notice that the nontrivial `interaction vertex' $\Zf$, \eqref{zcycle}, which eventually induces the Hochschild cocycle, is of the form of the cycle that is dual to this cocycle. Therefore, the proposed equations serve as a device to convert cycles, which are usually simple and easy to find, into cocycles, which are more complicated. In addition, the choice of a representative of the cohomology class is encoded in the gauge symmetries of the equations. It seems that the phenomenon is general enough and it would be interesting to understand its precise relation to the Shoikhet-Tsygan Formality. As we have already mentioned, from the mathematical point of view, the equations provide a resolution of the Hochschild complex written in the field-theoretical way.

\section{Conclusions and Discussion}
\label{sec:Conclusions}
In the paper we studied the problem of deforming higher-spin symmetries in the formal sense. It was shown that the first nontrivial vertex $\mathcal{V}$ that makes the higher-spin gauge connection $\omega$  non-flat
\besubeqs
\begin{align}
d\omega&=\omega\star \omega +\mathcal{V}(\omega,\omega,C)+...\,,\\
dC&=\omega\star C- C\star\tilde{ \omega}+...
\end{align}
\esubeqs
is determined by certain Hochschild two-cocycle $\Phi(a,b)$ of the higher-spin algebra: 
\begin{align}
    - a\star \Phi(b,c)+\Phi(a\star b,c)-\Phi(a,b\star c) +\Phi(a,b)\star \tilde{c}=0\,.
\end{align}
The vertex is $\mathcal{V}(\omega,\omega,C)=\Phi(\omega,\omega)\star \tilde C$. The Hochschild cocycle can be explicitly written down by employing the Shoikhet-Tsygan formality. Since the higher-spin fields backreact onto themselves the deformation problem does not stop at the Hochschild cocycle and higher orders are required. It seems that there are no obstructions at higher orders and hence reconstructing them is a routine procedure, which still needs to be recast in the language of formal structures. The upshot is that any formal higher-spin theory is completely determined by the Hochschild two-cocycle.

There are, however, cases where no nonlinear completion is needed and the Hochschild cocycle of the relevant higher-spin algebra is the only term beyond the higher-spin algebra structure constants. One example of physical importance is the propagation of higher-spin fields over background given by any flat connection.

There is a plenty of higher-spin theories that are expected to exist for the same reason as the simplest Type-A model does (its spectrum consists of totally-symmetric massless fields) and some tests of the AdS/CFT duality have already been done \cite{Beccaria:2014qea,Beccaria:2014zma,Beccaria:2014xda,Bae:2016hfy,Bae:2016xmv,Gunaydin:2016amv,Giombi:2016pvg} despite the lack of any formulation beyond free fields. The relevant higher-spin algebras are known and can always be associated with the quantization of the coadjoint orbits corresponding to the fundamental field of some free CFT. It is also plausible that the Hochschild two-cocycle is the only relevant cohomology and there are no obstructions at higher orders.

Thinking of the original problem of constructing higher-spin theories, it is also important to identify the structures that are responsible for locality in the field theory terms. The problem is that formal unfolded equations may give ill-defined differential equations upon identification of $d$ with $dx^\mm \pl_\mm$. The source of the problem is the appearance of nonlinearities in the zero-forms $C$ that contain derivatives of the fields of unbounded order. Such nonlinearities can be present as they encode interactions. The $C$-terms cannot be just consistent with the higher-spin symmetry in a formal sense, but must also be constrained by locality. The other side of the coin is that the freedom in coboundaries corresponds to field-redefinitions and those involving powers of $C$ can be arbitrarily non-local while still well-defined in the formal sense, e.g. such redefinitions are capable of washing away the stress-tensor \cite{Prokushkin:1999xq}, which is clearly unphysical.\footnote{It is worth mentioning that, as was proved in \cite{Prokushkin:1998bq}, the $3d$ Prokushkin-Vasiliev equations are formally empty: can be reduced to flat connection/covariant constancy equations. Therefore, the nontriviality of the $3d$ higher-spin theories is entirely due to appropriate locality constraints. } It is possible to show that the deformation problem not constrained by locality is to some extent empty \cite{Barnich:1993vg} in the field theory terms. It would be important to at least understand which part of higher-spin theories is captured by formal higher-spin theories where locality is not taken into account.

One of the main applications of higher-spin theories has been to understand AdS/CFT. Confining ourselves to formal structures, one can ask if there exists some way to relate formal higher-spin theories to the expected CFT duals. One can try to associate some invariants to the formal higher-spin structures and conjecture them to reproduce the CFT correlators, see e.g. \cite{Sezgin:2011hq,Colombo:2012jx,Vasiliev:2015mka}. For instance, at the zeroth-order in the expansion over a flat connection of the HS algebra the traces are the simplest invariants \cite{Colombo:2012jx,Didenko:2012tv} 
\begin{align}
 \langle j_1 j_2...\rangle&= tr \left(C_1\star \tilde{C}_2\star...\right) 
\end{align}
and can be shown to reproduce the free CFT correlation functions. These invariants do not correspond to any field theory observables in $AdS$ as they are non-local, but they show that the formal AdS/CFT duality can also make sense. It would be interesting to see if the invariants can be defined for the nonlinear unfolded equations, see also \cite{Sezgin:2011hq,Colombo:2012jx,Vasiliev:2015mka}.

The simplest examples of higher-spin algebras are related to the Weyl algebra $A_{n}$ in $2n$ generators. Another result is that we showed that the Hochschild cocycle for the Weyl algebra $A_{n}$ is related to a remarkable function of $2n+1$ vectors in the symplectic space. The function checks if the origin of the space belongs to the simplex built out of the $2n+1$ vectors and it is the Alexander-Spanier cocycle. The Hochschild cocycle condition then has a purely geometrical interpretation that a polytope made out of $2n+2$ vectors can be cut into $2n+2$ simplices. Also, the Hochschild cocycle can be generated by a number of simple steps by doubling the Weyl algebra from $y_\ga$ to $y_\ga,z_\ga$: the homotopy operator of the de Rham complex in the auxiliary $z$-space is followed by the star-product with one of the arguments of the cocycle and the procedure is repeated till all arguments are saturated.

The observation just discussed allows one to guess the equations that generate the Hochschild cocycle upon solving a chain of simple $\delta G= DG$ equations a la Fedosov. The nontriviality of the cocycle corresponds to the fact that certain `interaction term' cannot be redefined away, while the cocycle condition is a simple consequence of $DD\equiv0$. The interaction term has exactly the form of the cycle that is dual to the Hochschild cocycle. Therefore, the equations we found give a tool to convert cycles into cocycles at least for the case of the Weyl algebra. It would be interesting to study this problem in a more general setting.

We hope that the paper clarifies the relation between free conformal fields theories and dual formal higher-spin theories and can be useful for constructing more higher-spin theories. Also there is an interesting relation between higher-spin theories and the formality theorems that can be advantageous for both sides as the formality theorems give explicit formulae for the Hochschild cocycles and many other structures that can be of use in higher-spin theories, while it seems that the same structures can also be generated from certain simple higher-spin-like equations.

\section*{Acknowledgments}
\label{sec:Aknowledgements}
We would like to thank Thomas Basile, Xavier Bekaert, Roberto Bonezzi, Nicolas Boulanger, Vasily Dolgushev,  Boris Feigin, David De Filippi, Maxim Grigoriev, Dmitry Ponomarev, Ergin Sezgin and Per Sundell for the very useful discussions, explanations and comments. The work of E.S. was supported by the DFG Transregional Collaborative Research Centre TRR 33 and the DFG cluster of excellence ``Origin and Structure of the Universe". The work of A.Sh.
was supported in part by the RFBR grant No. 16-02-00284 A. E.S. and A.Sh. also acknowledge a kind hospitality at the program ``Higher Spin Theory and Duality" MIAPP, Munich (May 2-27, 2016) organized by the Munich Institute for Astro- and Particle Physics (MIAPP).

\begin{appendix}
\renewcommand{\thesection}{\Alph{section}}
\renewcommand{\theequation}{\Alph{section}.\arabic{equation}}
\setcounter{equation}{0}\setcounter{section}{0}

\section{Crash Course on the Vasiliev Equations}
\label{app:Vasiliev}
To facilitate comparison of the equations discussed in the paper with the $4d$ Vasiliev equations we review the latter below, see also \cite{Vasiliev:1999ba,Didenko:2014dwa}. First of all, it is easier to present what should be called the holomorphic truncation of the equations, see e.g. \cite{Iazeolla:2007wt} for some comments. The algebra is $A_1$ and is generated by $y_\ga$, where the indices $\ga,\gb,...$ run over two values. The doubled $(y,z)$ star-product is as in \eqref{yzstar}, but it is usually written in the integral form:\footnote{The normalization is such that $1\star f(y,z)=f(y,z)$ for any $f$.}
\begin{equation}\label{StarYZ}
(f\star g)(y,z)=\frac{1}{(2\pi)^2}\int d^2u d^2v f(y+u,z+u) g(y+v,z-v) e^{iu_\ga v^\ga}\,.
\end{equation}
Effectively, we have the following rules for the generating elements:
\begin{align}
y_\ga\star f(y,z)&=(y_\ga + i\pl^y_\ga
-i \pl^z_\ga)f(y,z)\,,
& z_\ga\star f(y,z)&=(z_\ga+i\partial^y_\ga-i\partial^z_\ga)f(y,z)\,, \\
f(y,z)\star y_\ga&=(y_\ga-i\partial^y_\ga-i\partial^z_\ga)f(y,z)\,, &
f(y,z)\star z_\ga&=(z_\ga+i\partial^y_\ga+i\partial^z_\ga)f(y,z)\,.
\end{align}
The field content is 
\begin{align}
\Wf&=\Wf_\mm(y,z|x)dx^\mm\,, & \Sf_\ga&=\Sf_\ga(y,z|x)\,, &\Bf&=\Bf(y,z|x)  \,.
\end{align}
The equations (half of the equations that make the holomorphic truncation) are
\besubeqs\label{VasAllHol}
\begin{align}
&d\Wf=\Wf\star \Wf\,,\\
&d(\Bf\star \klein)=[\Wf, \Bf\star \klein]_\star\,,\\
&d\Sf_{\ga}=[\Wf,\Sf_{\ga}]_\star \,,\\
&[\Sf_{\ga},
\Sf_{\gb}]_\star =-2i\epsilon_{\ga\gb}(1+\Bf\star \klein)\label{ssequation}\,,\\
&\{\Sf_{\ga}, \Bf\star \klein\}_{\star }=0\,,\label{sbequation}
\end{align}
\esubeqs
where $\varkappa$ is the celebrated Klein operator $\klein=e^{iz_\ga y^\ga}$. The Klein operator provides the inner realization of the twist automorphism, see \eqref{kleinchik}. The simplest vacuum solution is given by a flat connection of the HS algebra, vanishing zero-forms and $\Sf_\ga$ is customized to induce the de Rham differential in the $z$-space:
\begin{align}
\Wf&=\omega\,, &d\omega&=\omega\star \omega\,, &&\Bf=0\,, && \Sf_\ga=z_\ga \,.
\end{align}
The perturbation theory is essentially the same as discussed in Section \ref{sec:VasDouble}. The last two equations are equivalent to the defining relations of $osp(1|2)$ and $\Bf\star \varkappa$ can be get rid off, see e.g. \cite{Vasiliev:1999ba,Alkalaev:2014nsa}.

Let us also note that the peculiar Eqs.\eqref{ssequation}, \eqref{sbequation}, after we shift $\Sf$ by its vacuum value $\Sf_\ga dz^\ga\rightarrow z_\ga dz^\ga+2i \Sf_\ga dz^\ga$ and rescaling $\Bf\rightarrow 2i \Bf$, acquire the form of the non-zero curvature for $\Sf$ and the twisted-adjoint covariant derivative for $\Bf$:
\besubeqs
\begin{align}
\delta \Sf&= \Sf\star\Sf +\frac12 (\Bf\star \varkappa)dz_\nu \wedge dz^\nu\,,\\
\delta \Bf &= \Sf\star \Bf-\Bf\star \pi(\Sf)\equiv[\Sf,\Bf]_\pi\,.
\end{align}
\esubeqs
Altogether the equations can be written in the form of \eqref{holovas}.

The full $4d$ equations are obtained by proliferating the variables: $y_\ga, z_\ga$ are appended by $\bry_\gad,\brz_\gad$ that form the same star-product algebra. The field content is extended by the extra component $\Sf_\gad$, which is a connection along the new direction $\brz$. The additional (or to be replaced) equations are:
\besubeqs\label{VasAll}
\begin{align}
&d\bar{\Sf}_{\gad}=[W, \bar{\Sf}_{\gad}]_\star \,, &&\{\bar{\Sf}_{\gad}, \Bf\star \brklein\}_{\star }=0\,,\\
&[\Sf_{\ga},
\Sf_{\gb}]_\star =-2i\epsilon_{\ga\gb}(1+e^{i\theta}\Bf\star \klein)\,,
&&[\bar{\Sf}_{\gad},
\bar{\Sf}_{\gbd}]_\star =-2i\epsilon_{\gad\gbd}(1+e^{-i\theta}\Bf\star \brklein)\,,\label{BBbar}\\
&[\Sf_{\ga}, \bar{\Sf}_{\gad}]_\star =0\,,
\end{align}
\esubeqs
where $\brklein=e^{i\bar z_\gad \bar y^\gad}$ and, provided the reality conditions are taken into account, there is an option to introduce the phase $\theta$ (the $[\Sf_{\ga},
\Sf_{\gb}]_\star=...$ equation \eqref{ssequation} needs to be replaced with the one above).

There are some kinematical constraints that originate from the fact that both $\Bf\star \varkappa$ and $\Bf\star \brklein$ must be covariantly constant with respect to $\Wf$. They imply that $\klein\star \Bf\star \klein=\brklein\star \Bf\star \brklein$, i.e. $\Bf(-y,\bry,-z,\brz)=\Bf(y,-\bry,z,-\brz)$, and for the physical reasons one should impose the same on $\Wf$.

It is obvious that instead of doubling the variables one can introduce any number $M$ of families $y^i_\ga,z^j_\ga$, $i=1,...,M$ with the straightforward extension of the field content by $\Sf^i_\ga$. The case of $M=1$ corresponds to the holomorphic truncation and $M=2$ is selected for physical reasons that the finite-dimensional subalgebra of $z$-independent functions is $sp(4)\sim so(3,2)$. The equations for $M>2$ are formally consistent too. The holomorphic equations present the simplest instance and all $M>1$ cases correspond to $M$ independent deformations acting at the same time, the only interdependence being via kinematical constraints on the fields. 

It is interesting that the simplest gauge in the perturbation theory --- the Schwinger-Fock gauge --- yields the Hochschild cocycle that is not $sp(2)$-basic, see \cite{Boulanger:2015ova} for explicit formulas at the second order. The fact that part of the equations are the defining relations of $sp(2)$ (actually $osp(1|2)$) allows one to find a non-linear field redefinition that restores manifest $sp(2)$ covariance at higher orders \cite{Vasiliev:1999ba}.

\section{Hochschild, Cyclic and Lie Algebra Cohomology}
\label{app:Hochschild}

Here we collect some generalities on the cohomology theory of associative and Lie algebras, which are mentioned in the body of the paper. For a more comprehensive and systematic exposition of the subject we refer the reader to the books
\cite{MacLane}, \cite{Loday}, \cite{Khalkhali}.

Let $A$ be a complex  associative algebra with unit $1$. Recall that a bimodule  $M$ over $A$ is a complex vector space equipped with commuting left and right actions of $A$: 
\begin{equation}
m\mapsto amb \qquad \forall m\in M, \quad \forall a,b\in A\,.
\end{equation}
The {\it Hochschild cohomology} $HH^\bullet(A,M)$
of the algebra $A$ with coefficients in $M$ is the cohomology of the Hochschild cochain complex 
\begin{equation}
C^\bullet(A,M) :\quad  C^0\stackrel{\partial}{\longrightarrow} C^1\stackrel{\partial}{\longrightarrow} C^2\stackrel{\partial}{\longrightarrow} \cdots
\end{equation}
with 
\begin{equation}
C^p=\mathrm{Hom}_{\mathbb{C}}(A^{\otimes p}, M)\,,\qquad A^{\otimes p}=\underbrace{A\otimes \cdots \otimes A}_p\,, 
\end{equation}
and the differential 
\begin{equation}\label{d-H}
(\partial f)(a_1,\ldots, a_{p+1})=a_1 f(a_2,\ldots,a_{p+1})+\sum_{k=1}^{p}(-1)^{k+1} f(a_1,\ldots,a_ka_{k+1},\ldots, a_{p+1})
\end{equation}
$$
+(-1)^{p+1}f(a_1,\ldots,a_{p})a_{p+1}\,.
$$
The Hochschild complex $C^
\bullet(A,M)$ contains a large subcomplex $\bar C^\bullet(A,M)$ of cochains that vanish when at least one of their arguments is equal to $1$. The latter is called the {\it normalized Hochschild complex}. It is easy to see that the inclusion map $i: \bar C(A,M)\rightarrow C(A,M)$ induces an isomorphism in cohomology. This means that the Hochschild cohomology of $A$ is isomorphic to that of the quotient  algebra  $\bar A=A/\mathbb{C}1$. 

Of particular interest is the bimodule $M=A^\ast$, with $A^\ast$ being linear dual to the space $A$. Unlike the general case, the groups $HH^\bullet(A,A^\ast)$ are functors of the algebra $A$, meaning that for any homomorphism of algebras $h: A\rightarrow B$ we have the homomorphism $h^\ast: HH^\bullet(B, B^\ast)\rightarrow HH^\bullet(A, A^\ast)$ in cohomology. For this reason their notation is abbreviated  to $HH^\bullet(A)$.  The $A$-bimodule structure on $A^\ast$ is given by $af(c)b=f(acb)$ for all $a,b, c\in A$, $f\in A^\ast$ and the bimodule of $p$-cochains is naturally identified with the space $\mathrm{Hom}_{\mathbb{C}}(A^{\otimes p+1},\mathbb{C})$. In other words, each $p$-cochain is given by a $\mathbb{C}$-linear map $\varphi: A^{\otimes p+1}\rightarrow \mathbb{C}$ interpreted as 
\begin{equation}
\varphi(a_0,a_1,\ldots,a_p)=f(a_1,\ldots a_p)(a_0)\,.
\end{equation}
The action of the Hochschild differential (\ref{d-H}) takes now the form 
\begin{equation}
(\partial \varphi)(a_0, a_1,\ldots, a_{p+1})=\sum_{k=0}^p(-1)^k\varphi(a_0,\ldots,a_ka_{k+1},\ldots,a_{p+1}) +(-1)^{p+1}\varphi(a_{p+1}a_0,a_1,\ldots,a_p)\,.
\end{equation}
A remarkable fact, discoved by A. Connes \cite{Connes:85}, is that the Hochschild complex $C^\bullet(A,A^\ast)$ contains an interesting subcomplex composed of the so-called {\it cyclic cochains}. A $p$-cochain $\varphi$ is called cyclic if 
\begin{equation}
\varphi(a_p,a_0,a_1,\ldots,a_{p-1})=(-1)^p\varphi (a_0,a_1,\ldots,a_{p})\,.
\end{equation}
The cohomology groups of this  subcomplex are denoted by $HC^\bullet(A)$, where the letter $C$ refers  either to {\it cohomologie cyclique} or {\it cohomologie de Connes}.  Since its appearance in the eighties, the cyclic cohomology theory has attracted both  mathematicians and physicists  due to its fundamental role in noncommutative geometry \cite{Connes:95}, \cite{Khalkhali}. 
In many practical cases the cyclic cohomology groups can be computed from the  Hochschild cohomology due to Connes' long exact sequence:
\begin{equation}\label{CES}
\cdots \rightarrow HC^p(A)\stackrel{I}{\rightarrow} HH^p(A)\stackrel{B}{\rightarrow} HC^{p-1}(A)\stackrel{S}{\rightarrow} HC^{p+1}(A)\rightarrow \cdots
\end{equation}
See \cite{Connes:95}, \cite{Loday} for the definition of the operators $I$, $B$ and $S$. 

Let ${\mathcal{M}}_r(\mathbb{C})$ denote the algebra of complex $r\times r$-matrices. Tensoring this algebra with an associative algebra $A$, we get the matrix algebra $\mathcal{M}_r(A)=A\otimes \mathcal{M}_r(\mathbb{C})$, i.e., the algebra of matrices with  entries in $A$. Similarly, tensoring an $A$-bimodule $M$ with $\mathcal{M}_r(\mathbb{C})$ yields the $\mathcal{M}_r(A)$-bimodule $\mathcal{M}_r(M)=M\otimes \mathcal{M}_r(\mathbb{C})$. It turns out that  the functor  $\otimes \mathcal{M}_r(\mathbb{C})$ affects neither the Hochschild nor the cyclic cohomology, namely, 
\begin{equation}\label{Minv}
HH^\bullet (A,M)\simeq HH^\bullet (\mathcal{M}_r(A),\mathcal{M}_r(M))\,,\qquad HC^\bullet (A)\simeq HC^\bullet (\mathcal{M}_r(A))\,.
\end{equation}
At the level of cochains the above isomorphisms are induced by the so-called {\it cotrace map}. For the Hochschild complex it is given by 
\begin{equation}\label{cotr}
\mathrm {cotr}: C^p(A,M)\rightarrow C^p (\mathcal{M}_r(A),\mathcal{M}_r(M))\,,
\end{equation}
$$f(a_1,\ldots, a_p)\mapsto F(a_1\otimes u_1, \ldots, a_p\otimes u_p)=f(a_1,\ldots, a_p)u_1\cdots u_p\,,\qquad a_i\in A\,,\quad u_i\in \mathcal{M}_r(\mathbb{C})\,,
$$
and for the cyclic cochains we have 
\begin{equation}\label{cotr-c}
\varphi(a_0, a_1,\ldots, a_p)\mapsto \Phi(a_0\otimes u_0, a_1\otimes u_1, \ldots, a_p\otimes u_p)=f(a_1,\ldots, a_p)\mathrm{Tr} (u_0u_1\cdots u_p)\,.
\end{equation}
The homotopically inverse to the cotrace map is induced  by the natural inclusions 
\begin{equation}\label{inc}
\mathrm{inc}: A\simeq \mathcal{M}_1(A)\rightarrow \mathcal{M}_r(A)\,,\qquad \mathrm{inc}: M\simeq \mathcal{M}_1(M)\rightarrow \mathcal{M}_r(M)\,.
\end{equation}
The isomorphisms (\ref{Minv}) are known as the {\it Morita invariance} of the Hochschild and cyclic cohomology. 

Consider now a complex Lie algebra  $L$ and let $M$ be an $L$-module. We write $
[m,x]$ for the right action of an element $x\in L$ on $m\in M$. Such a module is also called a representation of the Lie algebra $L$. 
Denote by $\Lambda^\bullet L$ the exterior algebra of the space $L$. Then
the Chevalley-Eilenberg cochain complex  is the sequence of homomorphisms 
\begin{equation}
C^\bullet(L,M) :\quad  C^0\stackrel{\delta}{\longrightarrow} C^1\stackrel{\delta}{\longrightarrow} C^2\stackrel{\delta}{\longrightarrow} \cdots\,,
\end{equation}
where 
\begin{equation}
C^p=\mathrm{Hom}_{\mathbb{C}}(\Lambda^p L, M)\,,\qquad \Lambda^p L=\underbrace{L\wedge \cdots \wedge L}_p\,, 
\end{equation}
and the differential is given by 
\begin{equation}
\begin{array}{l}
(\delta f)(x_1\wedge \ldots\wedge  x_{p+1})\displaystyle =\sum_{k=1}^{p+1}(-1)^{k}[f(x_1\wedge \cdots\wedge \hat{x}_k\wedge \cdots\wedge x_{p+1}), x_k]\\[7mm]
\displaystyle
+
\sum_{1\leq k\leq l\leq p+1}(-1)^{k+l-1} f([x_k,x_l]\wedge \cdots\wedge \hat{x}_k\wedge \cdots\wedge \hat{x}_{l}\wedge\cdots\wedge x_p)\,.
\end{array}
\end{equation}
As usual $\hat{x}_i$ means that the argument $x_i$ has been omitted. The corresponding cohomology groups are denoted by $H^p(L,M)$ and called the {\it Lie algebra cohomology} groups. 

Any $A$-bimodule $M$ can be viewed as a right  module over the associated Lie algebra $L(A)$ if we set  
\begin{equation}
[m,a]=ma-am \qquad \forall a\in L(A)\,,\quad\forall m\in M\,.
\end{equation}
This allows one to relate the Hochschild cohomology of $A$ with the Lie algebra cohomology of $L(A)$. The relation is established by the {\it antisymmetrisation map}
\begin{equation}
\varepsilon : C^\bullet (A,M)\rightarrow C^\bullet (L(A), M)\,,
\end{equation}
$$
(\varepsilon f) (a_1\wedge \cdots\wedge a_p)=\sum_{\sigma \in S_p} \mathrm{sgn}(\sigma) f(a_{\sigma(1)},\ldots,a_{\sigma(p)})\,.
$$
It is easy to check that $\delta \varepsilon =\varepsilon \partial$; hence, $\varepsilon$ is a cochain map inducing a homomorphism in cohomologies:
\begin{equation}
\varepsilon^\ast: HH^\bullet (A,M)\rightarrow H^\bullet (L(A), M)\,.
\end{equation}
In general, this homomorphism is neither injective nor surjective. 

A more definite relationship between the Hochschild and Lie algebra cohomologies can be made for the matrix Lie algebras.  By the matrix Lie algebra $gl_r(A)$ we mean the Lie algebra associated to the algebra $\mathcal{M}_r(A)$.  The Lie bracket in $gl_r(A)$ is just  the matrix commutator. If $M$ is a bimodule over $A$, then $\mathcal{M}_r(M)$ is the natural bimodule over $\mathcal{M}_r(A)$ and the adjoint module over $gl_r(A)$. 

One of the central results of the cyclic cohomology theory is the following 

\vspace{3mm}
\noindent 
{\bf Theorem 1.} {\it For $r$ large enough, there exist isomorphisms}
$$
H^\bullet(gl_r(A),\mathcal{M}_r(M))\simeq HH^\bullet (A,M)\otimes H^\bullet(gl_r(A),\mathbb{C})\,,\qquad H^\bullet(gl_r(A),\mathbb{C})\simeq 
\Lambda^\bullet(HC^{\bullet-1}(A))\,.
$$

As a result the groups  $H^\bullet(gl_r(A),\mathcal{M}_r(A))$ are computable from $HH^\bullet (A,M)$ and $HC^\bullet(A)$ (no matrices enter). 
Moreover, the computation of the cyclic cohomology $HC^\bullet(A)$ can further be reduced to the Hochschild cohomology
$HH^\bullet(A)$ via the Connes exact sequence (\ref{CES}). The homological versions of the above isomorphisms for trivial coefficients were established independently and simultaneously by  Loday-Quillen \cite{Quillen} and Tsygan \cite{TsyganOther}. Generalization to the adjoint representation was given by Goodwillie \cite{Goodwillie}. For the proof of  dual isomorphisms in cohomologies see \cite{Loday}, \cite{Brodzki}. 

Applying the theorem above to the Weyl algebra one can deduce the following  

\vspace{3mm}
\noindent 
{\bf Theorem 2.} {\it For $r\gg2n$},
$$
H^{2n}(gl_r(A_n), gl_r(A_n)^\ast)\simeq \mathbb{C}\,,\qquad H^{p}(gl_r(A_n), gl_r(A_n)^\ast)=0 \qquad \forall p<2n\,.
$$
See \cite{FeiginTsygan} for the proof.

\section{Weyl Algebra and Around}
\label{app:Weyl}
Let $V$ be a $2n$-dimensional symplectic space over $\mathbb{C}$ with the Cartesian coordinates $y^\alpha$ and the dual symplectic form $C^{\alpha \beta}=C(y^\alpha,y^\beta)$.
Associated to $V$ is the Weyl algebra $A_n$ defined as an associative, unital algebra over $\mathbb{C}$ generated by $2n$ variables $y^\alpha$ subject to the relations 
$
y^\alpha y^\beta-y^\beta y^\alpha=2iC^{\alpha\beta }
$.

Alternatively, one can define the Weyl algebra as the space of complex polynomials $\mathbb{C}[y^1,\ldots, y^{2n}]$ endowed with the Moyal-Weyl $\star$-product:
\begin{equation}\label{WM}
(a\star b)(y)=\exp\left( iC^{\alpha\beta}\frac{\partial}{\partial y^\alpha}\frac{\partial}{\partial z^\beta}\right) a(y)b(z)|_{z=y}\,,\qquad \forall a,b\in \mathbb{C}[y^1,\ldots, y^{2n}]\,.
\end{equation}
Passing  to another set of generators, if necessary, it is possible to bring the matrix  $C=(C^{\alpha\beta})$ into the block-diagonal form
\begin{equation}\label{CanF}
C=\left(\begin{array}{cccccc}0&1&&&&\\-1&0&&&&\\&&\ddots\\&&&0&1&\\&&&-1&0&\\ \end{array}\right)\,.
\end{equation}
Now it becomes clear that the algebra $A_{n}$ is isomorphic to $n$th tensor power of $A_1$. In what follows we will assume that the matrix  $C$ has the canonical form (\ref{CanF}).

As usual, the algebra $A_n$ can be viewed as a bimodule over itself:
\begin{equation}
c\mapsto a\star c\star b\qquad \forall a,b,c\in A_n\,.
\end{equation}
This bimodule structure extends naturally from the space $A_n$ to its completion $\hat A_n=\mathbb{C}[[y^1,\ldots,y^{2n}]]$. The latter is given by the formal power series in $y$'s with complex coefficients. (Notice that the $\star$-product of two elements of $\hat{A}$
is ill-defined.) 

The group $Sp(2n,\mathbb{C})$, acting in $V$ by  linear transformations preserving the symplectic form, defines  a subgroup in the group $\mathrm{Aut}(A_n)$ of automorphisms of the Weyl algebra $A_n$.  
Let  $G$ be a finite subgroup of  $Sp(2n,\mathbb{C})$ and denote by $a^g$ the action of $g\in\mathrm{ Aut}(A_n)$ on the element $a\in A_n$. Then we can define the smash product 
$A_n\rtimes G$ of the Weyl algebra $A_n$ and the group algebra $\mathbb{C}[G]$. As a  vector space $A_n\rtimes G$ is given by the tensor product $A_n\otimes \mathbb{C}[G]$ and multiplication is given by
\begin{equation}
(a\otimes g)(b\otimes h)=a\star b^g\otimes gh
\end{equation}
for all $a,b\in A_n$ and $g,h\in G$. Many higher-spin algebras of physical interest  can be defined as smash products. For example, the higher-spin algebra underlying four-dimensional field theories with $N=2$ supersymmetry \cite{Vasiliev:1986qx} is given by the smash product $A_2\rtimes G$ with
\begin{equation}\label{G5}
G=\mathbb{Z}_2\times \mathbb{Z}_2\subset Sp(1)\times Sp(1)\subset Sp(2)\,.
\end{equation}
More explicitly, the group $G$ is generated by the pair of commuting symplectic reflections $g_1, g_2\in G$, whose action on the canonical generators is given by
\begin{equation}\label{G6}
\begin{array}{llll}
(y^1)^{g_1}=-y^1\,,\qquad &(y^2)^{g_1}=-y^2\,,\qquad &(y^3)^{g_1}=y^3\,, \qquad &(y^4)^{g_1}=y^4\,,\\[5mm]
(y^1)^{g_2}=y^1\,,\qquad &(y^2)^{g_2}=y^2\,,\qquad &(y^3)^{g_2}=-y^3\,, \qquad &(y^4)^{g_2}=-y^4\,.
\end{array}
\end{equation}

Associated to the Weyl algebra $A_n$ is the Lie algebra $L(A_n)$, with the Lie bracket given by the commutator $[a,b]=a\star b-b\star a$. The assignment 
\begin{equation}
L(A_n)\ni a\mapsto ad_ab=[a,b]\qquad \forall b \in A_n
\end{equation}
defines the adjoint representation of the Lie algebra $L(A_n)$ in the space $A_n$. Given an element $g\in \mathrm{Aut}(A_n)$, one can define the so-called {\it twisted-adjoint } representation by setting 
\begin{equation}
L(A_n)\ni a\mapsto ad^g_ab=a\star b-b\star a^g\qquad \forall b \in A_n\,.
\end{equation}
We will refer to $[a,b]_g=a\star b-b\star a^g$ as the $g$-commutator. The usual commutator corresponds to $g=e$. 
The twisted-adjoint representation extends  in the natural way from the space $A_n$ to its completion $\hat{A}_n$.

The $\star$-product defined by Eq. (\ref{WM}) corresponds to the so-called Weyl (or symmetric) ordering of $y$'s. There are, of course, many other ways to order the generators in monomials  and all of them can uniformly be described by means of a symmetric form $G^{\alpha\beta}=G(y^\alpha, y^\beta)$ on  $V^\ast$, see e.g. \cite{Dolgushev:2001ha}.  For this end, one needs only to replace the matrix $C^{\alpha\beta}$ in (\ref{WM}) by the sum $C^{\alpha\beta}+G^{\alpha\beta}$. Let us denote the resulting product by $\star_G$. 
It is not hard to check directly the associativity of the $\star_G$-product, but this also follows from a more stronger statement: The algebras $(A_n,\star)$ and $(A_n,\star_G)$ are isomorphic to each other.  The isomorphism is established by the invertible pseudo-differential operator
$
U: A_n\rightarrow A_n
$  defined by 
\begin{equation}
U=e^{-i\Delta}\,, \qquad \Delta=G^{\alpha\beta}\frac{\partial^2}{\partial y^\alpha\partial y^\beta}\,,\qquad U^{-1}=e^{i\Delta}\,.
\end{equation}
One can see that 
\begin{equation}
a\star_G b=U^{-1}((Ua)\star U(b))\qquad \forall a,b \in A_n\,.
\end{equation}

It should be stressed that this isomorphism of the associative algebras $(A_n,\star)$ and $(A_n\star_G)$ does not extend to the isomorphism
of the corresponding bimodules $\hat A_n$.  The reason is obvious: the action of the pseudo-differential operator $U$ is ill-defined in the space of formal power series in $y$'s. So, one should be careful with reorderings of generators  when elements of the space $\hat{A}_n$ are involved into the game. In the context of higher-spin theories, for example,  the non-polynomial functions of $y$'s appear usually under the name of {\it inner Klein operators} \cite{Vasiliev:1988sa}.  The presence of these operators makes it difficult to pass to the Weyl ordering of all the oscillator variables in Vasiliev's equation \cite{Vasiliev:1992av} unless more general functions on the Weyl algebra are introduced, see \cite{Iazeolla:2011cb,Sundell:2016mxc} for examples.

The Hochschild cohomology groups of the Weyl algebra as well as its smash products have been computed for various coefficients by making use of the the Koszul resolution \cite{Kassel}, \cite{Pinczon}.  The existence of the Koszul resolution implies, among other things, that $HH^p(A_n,M)=0$ for any $M$ and $p>2n$. Let us also mention the following two results. 

\vspace{3mm}
\noindent
\textbf{Theorem 1.} {\it The cohomology space $HH^p(A_n\rtimes G,A_n\rtimes G)$ is naturally isomorphic to the space of conjugation invariant functions on the set $S_p$ of elements $g\in G$ such that} $$\mathrm{rank} (1-g)|_V=p\,.$$

Since $\mathrm{Im}(1-g)$ is a symplectic vector space it follows immediately that the odd cohomology of  $A_n\rtimes G$ with coefficients in itself vanishes.

For any given element $g\in Sp(n,\mathbb{C})$ denote by $A_ng$ the $A_n$-bimodule such that
\begin{equation}
c\mapsto a\star c\star b^g\qquad \forall a,b,c\in A_n
\end{equation}
(the so-called {\it twisted action} of $A_n$ on itself). Then we have 

\vspace{3mm}
\noindent
\textbf{Theorem 2.} {\it If $\mathrm{rank}(1-g)=q$, then  
$
HH^q(A_n,A_ng)\simeq \mathbb{C} 
$ 
and 
$HH^p(A_n,A_ng)=0$ for $p\neq q$.
}

\vspace{3mm}
The proofs can be found in \cite{AFLS} (see also \cite{Pinczon}). In  view of the Morita invariance (\ref{Minv}) the above theorems hold true if one replaces the Weyl algebra by its matrix extension $\mathcal{M}_r(A_n)$.

By way of illustration consider the higher-spin algebra $A=A_2\rtimes G$ defined by Rels. (\ref{G5}), (\ref{G6}).  
The group $G=\{e, g_1, g_2,g_1g_2\}$ being abelian, 
\begin{equation}
S_0=\{e\}\,, \qquad S_1=\{\varnothing\}\,, \qquad S_2=\{g_1, g_2\}\,,\qquad S_3=\{\varnothing\}\,,\qquad S_4=\{g_1g_2\}\,,
\end{equation}
and by Theorem 1 
$$
HH^0(A,A)\simeq \mathbb{C}\,,\quad HH^1(A,A)=0\,,\quad HH^2(A,A)\simeq \mathbb{C}^2\,,\quad HH^3(A,A)=0\,,\quad HH^4(A,A)\simeq \mathbb{C}\,.
$$

The equality $\dim HH^2(A,A)=2$ proves that the two linearly independent deformations $\mathcal{V}(\omega,\omega,C)$ found in \cite{Vasiliev:1988sa}, see \eqref{fourdvert}, exhaust all possibilities. The vanishing of the third group seems to indicate the absence of  higher-order obstructions for a consistent interaction.  
It is not hard to guess the structure of cocycles representing nontrivial cohomology classes: 
$$
1, \qquad \beta_1\beta_2\Phi_2(\alpha_1,\alpha_2)g_1, \qquad \alpha_1\alpha_2\Phi_2(\beta_1,\beta_2)g_2,,\qquad  \Phi_4(a_1,a_2,a_3,a_4)g_1g_2\,.
$$
Here we interpret $A_2$ as $A_1\otimes A_1$ and represent each element $a\in A_2$ as a linear combination of monomials $a=\alpha(y^1,y^2)\beta(y^3,y^4)$ with coefficients in $\mathbb{C}[G]$; $\Phi_n$ is the standard FFS cocycle for $A_n$.

\section{Equations for the Vertex}
\label{app:Vertex}
The equation for the Hochschild cocycle are discussed in the main text. In this Appendix we would like to reduce it to a functional equation for the symbols of operators.  

\subsection{Symbol Calculus}
\label{app:symbols}
The Moyal-Weyl star-product can be defined as 
\begin{align}
(f\star g)(y) &= \exp i[-iy^\nu(\pl_1+\pl_2)_\nu +(\pl_1)_\nu (\pl_2)^\nu]f(y_1)g(y_2)\Big|_{y_i=0}=\\
&=\exp i [p_0\cdot p_1+p_0\cdot p_2 +p_1\cdot p_2] f(y_1)g(y_2)\,,
\end{align}
where we introduced the following notation: $p_0=iy$, $p_1=\pl_1$, ..., $p_n=\pl_n$, $p_{ij}=p_i \cdot p_j= p_{i\alpha} p_{j\beta} C^{\alpha\beta}$. In practice, it is also important to express the star-product with a twisted element $\tilde{g}(y)=g(-y)$ of the Weyl algebra in terms of $g(y)$: 
\begin{align}
f\star \tilde g&= \exp i [p_0\cdot p_1-p_0\cdot p_2 -p_1\cdot p_2] f(y_1)g(y_2)
\end{align}
and we systematically omit $|_{y_i=0}$ at the end of the formulae. Mnemonically the rule is: see the twist --- reverse the sign of the corresponding $p$. The operators acting on $n$ functions can be understood as functions of $p_i$:
\begin{align}
V(a_1,...,a_n)=v(iy,\pl_1,...,\pl_2)a_1(y_1)...a_n(y_n)\Big|_{y_i=0}\,.
\end{align}
The dictionary between symbols of the operators and various ways of adding one more argument via taking star-product is:
\begin{align*}
a_1\star V(a_2,...,a_{n+1})&\rightarrow  v(p_0+p_1,p_2,...,p_{n+1})e^{+i p_{0}\cdot p_1}\,,\\
V(a_1,...,a_n)\star a_{n+1}&\rightarrow v(p_0-p_{n+1},p_1,...,p_n)e^{+i p_{0}\cdot p_{n+1}}\,,\\
V(a_1,...,a_n)\star \tilde{a}_{n+1}&\rightarrow v(p_0+p_{n+1},p_1,...,p_n) e^{-i p_{0}\cdot p_{n+1}}\,,\\
V(a_1,...,a_k\star a_{k+1},...,a_{n+1})&\rightarrow v(p_0,...,p_{k-1},p_k+p_{k+1},p_{k+2},...,p_{n+1})e^{+i p_{k}\cdot p_{k+1}}\,,\\
V(a_1,...,a_k\star \tilde{a}_{k+1},...,a_{n+1})&\rightarrow v(p_0,...,p_{k-1},p_k-p_{k+1},p_{k+2},...,p_{n+1}) e^{-i p_{k}\cdot p_{k+1}}\,,\\
a_1\star \tilde{V}(a_2,...,a_n)&\rightarrow v(-p_0-p_1,p_2,..,p_{n+1}) e^{+i p_{0}\cdot p_{1}}\,.
\end{align*}
Also it is important that
\begin{align}
\tilde{V}(a_1,...,a_n)&=V(\tilde{a}_1,...,\tilde{a}_n)
\end{align}
unless $V$ breaks $sp(2n)$. It is sometimes convenient to remove the twist from the argument:
\begin{align}
V(a_1,...,\tilde{a}_k,...,a_n)\rightarrow v(p_0,...,-p_k,...,p_n)\,.
\end{align}

\subsection{Vertex}
\label{app:vertex}
With the help of the dictionary above the equations for the first vertex can be written as:\footnote{As different from \cite{Vasiliev:1988sa} we do not factorize the star-product out of $v$. }
{\allowdisplaybreaks
\begin{align*}
0&=-v_1(p_0+p_1,p_2,p_3,p_4)e^{ip_{01}}+v_1(p_0,p_1+p_2,p_3,p_4)e^{ip_{12}}\\&-v_1(p_0,p_1,p_2+p_3,p_4)e^{ip_{23}}+v_1(p_0,p_1,p_2,p_3+p_4)e^{ip_{34}}\,,\\
0&=v_1(p_0-p_4,p_1,p_2,p_3)e^{ip_{04}}-v_1(p_0,p_1,p_2,p_3-p_4)e^{-ip_{34}}+\\
&+v_2(p_0,p_1+p_2,p_3,p_4)e^{ip_{12}}-v_2(p_0+p_1,p_2,p_3,p_4)e^{ip_{01}}-v_2(p_0,p_1,p_2+p_3,p_4)e^{ip_{23}}\,,\\
0&=v_2(p_0-p_4,p_1,p_2,p_3)e^{ip_{04}}-v_2(p_0,p_1,p_2,p_3+p_4)e^{ip_{34}}+v_2(p_0,p_1,p_2-p_3,p_4)e^{-ip_{23}}\\
&+v_3(p_0,p_1+p_2,p_3,p_4)e^{ip_{12}}-v_3(p_0+p_1,p_2,p_3,p_4)e^{ip_{01}}\,,\\
0&=v_3(p_0-p_4,p_1,p_2,p_3)e^{ip_{04}}-v_3(p_0,p_1,p_2,p_3+p_4)e^{ip_{34}}+\\
&+v_3(p_0,p_1,p_2+p_3,p_4)e^{ip_{23}}-v_3(p_0,p_1-p_2,p_3,p_4)e^{-ip_{12}}\,.
\end{align*}}
The field-redefinitions result in
\begin{align*}
    \delta v_{1}(p_0,p_1,p_2,p_3)&=g_1(p_0+p_1,p_2,p_3)e^{ip_{01}}-g_1(p_0,p_1+p_2,p_3)e^{ip_{12}}+g_1(p_0,p_1,p_2+p_3)e^{ip_{23}}\,,\\
    \delta v_{2 }(p_0,p_1,p_2,p_3)&=g_2(p_0+p_1,p_2,p_3)e^{ip_{01}}-g_2(p_0,p_1+p_2,p_3)e^{ip_{12}}+\\&+g_1(p_0-p_3,p_1,p_2)e^{ip_{03}}-g_1(p_0,p_1,p_2-p_3)e^{-ip_{23}}\,,\\
    \delta v_{3 }(p_0,p_1,p_2,p_3)&=-g_2(p_0,p_1,p_2+p_3)e^{ip_{23}}+g_2(p_0-p_3,p_1,p_2)e^{ip_{03}}+g_2(p_0,p_1-p_2,p_3)e^{-ip_{12}}\,.
\end{align*}
Choosing the left ordering once and for all, i.e. $V_2=V_3=0$, we end up with 
\besubeqs\label{vveqs}
\begin{align}\label{vveqsA}
0&=-v_1(p_0+p_1,p_2,p_3,p_4)e^{ip_{01}}+v_1(p_0,p_1+p_2,p_3,p_4)e^{ip_{12}}\\&-v_1(p_0,p_1,p_2+p_3,p_4)e^{ip_{23}}+v_1(p_0,p_1,p_2,p_3+p_4)e^{ip_{34}}\,,\notag\\
0&=v_1(p_0-p_4,p_1,p_2,p_3)e^{ip_{04}}-v_1(p_0,p_1,p_2,p_3-p_4)e^{-ip_{34}}\,.\label{vveqsB}
\end{align}
\esubeqs
These equations are equivalent to the Hochschild cocycle condition, as shown in the main text. The second equation is easy to solve.

\section{Simple Solution of the Hochschild Problem}
\label{app:lame}
It turns out that the case of the $A_1$ Hochschild cocycle is relatively simple and does not require anything beyond elementary mathematics. The equation for the cocycle
\begin{align}
   - a\star \Phi(b,c)+\Phi(a\star b,c)-\Phi(a,b\star c) +\Phi(a,b)\star \tilde{c}=0
\end{align}
should be rewritten in the language of generating functions, which gives
\begin{align*}
    -\hat\Phi(p_0+p_1,p_2,p_3)e^{ip_0\cdot p_1}&+\hat\Phi(p_0,p_1+p_2,p_3)e^{ip_1\cdot p_2}  -\hat\Phi(p_0,p_1,p_2+p_3)e^{ip_2\cdot p_3}+\hat\Phi(p_0+p_3,p_1,p_2)e^{-ip_0\cdot p_3}=0
\end{align*}
The same equation is obtained by plugging the solution of \eqref{vveqsB} into \eqref{vveqsA}. The only challenge to solve the equation comes from the last term where one has a twisted element of the algebra, $\tilde{c}$. Had it not been for the twist a simple solution would have been to take $\Phi(a,b)=a\star b$, whose generating function is
\begin{align}
\hat\Phi(p_0,p_1,p_2)&= e^{i[p_0\cdot p_1+p_0\cdot p_2+p_1\cdot p_2]}
\end{align}
and see that all the four terms cancel pairwise. In terms of generating functions this means that the shifts of the arguments get appropriately compensated by the exponential factors. There is no cohomology in this case, but it is important to realize that the ansatz for solution should contain some exponentials that are linear in the scalar products $p_i\cdot p_j$, otherwise there is no way to have cancellation between the terms in the equation. The pure star product ansatz is unable to solve the right equation because of the very last term: while the first two still cancel each other, the third one remains unpaired because the fourth term yields a completely different exponential due to the twist. Therefore, one has to add one more exponential to the ansatz that would compensate the fourth term. This is possible, but the cancellation is not perfect as one more exponential is generated. One needs to add the third exponential to the ansatz, which closes the chain: there are three exponentials in the ansatz and they generate four different exponentials via equations and should cancel each other in triplets. It is important that the form of the exponentials is thus fixed. The cancellation in triplets cannot be achieved with constant coefficients. Also, any manifestly smooth solution has good chances to be a coboundary. These two arguments makes one think of the Plucker identities:
\begin{align}
(p_{0}\cdot p_1) (p_{2}\cdot p_3) +(p_{1}\cdot p_2)(p_{0}\cdot p_3)-(p_{0}\cdot p_2)( p_{1}\cdot p_3)\equiv0\,,
\end{align}
which express a simple fact that it is impossible to have more than two linearly independent vectors in two dimensions. Let us assume that the Hochschild cocycle has the form of the three already found exponentials multiplied by a number of fractions:
\begin{align}
\hat\Phi(...)\sim \sum_i \frac{N}{D_i}e^{a_i}\,,
\end{align}
where $N$ is the common factor. In order to make the Plucker identities effective one has to make the denominators be the product of two factors that are linear in the scalar products:
\begin{align}
\hat\Phi(...)\sim  \frac{N}{d_1 d_2 }e^{a_1}+\frac{N}{d_2 d_3 }e^{a_2}+\frac{N}{d_3 d_1 }e^{a_3}\,.
\end{align}
Then it turns out that the denominators are completely fixed by the requirement for $\hat\Phi$ to be regular. Therefore, the only problem that remains is to fix the nominator $N$, which can be found by looking at the Taylor expansion of the cocycle equation, $N=p_1\cdot p_2$. It is then easy to see that $N$ is a unique numerator that solves all the four equations. The fact that the higher terms in the Taylor expansion of $\hat\Phi$ cannot affect the lower ones makes it easy to see that the solution is a cohomology. The final form of the solution is:
\begin{align}
\hat\Phi(p_0,p_1,p_2)&=z\Biggr[\frac{e^{i[x+y+z]}}{(x+y) (y+z)}-\frac{e^{i[x-y-z]}}{(x-z) (y+z)}+\frac{e^{i[-x-y+z]}}{(x+y) (x-z)}\Biggr]\,,
\end{align}
where $x=p_0\cdot p_1, y=p_0\cdot p_2,z=p_1\cdot p_2$. It coincides with the FFS cocycle \eqref{ffscocycle}, where the integrand for the $A_1$ case is  $\exp{i[x(1-2u_1)+y(1-2u_2)+z(1+2u_1-2u_2)]}$.

If we need to find a higher-order cocycle, for example for $sp(4)$, it is again clear that one should have four factors in denominators and five exponentials to close the chain of equations, which is due to the Plucker identities being of higher order, see Appendix \ref{app:spfour}.

\paragraph{Another approach.} In the seminal paper \cite{Vasiliev:1988sa} where the foundations of the unfolded approach to HS theories were laid done, a slightly different strategy was used. First of all, one can look at the set of two equations \eqref{vveqs} resulting after setting $V_{2,3}=0$. The idea of \cite{Vasiliev:1988sa} is to look for a seemingly trivial cocycle that comes from a singular coboundary, though the cocycle itself must be regular. The fact that it is represented as a formal coboundary ensures that it is closed, while the singularity of its coboundary representation guarantees that it cannot be exact in the class of regular functions. The singular coboundaries that do the job are
\begin{align}
\hat g_1(p_0,p_1,p_2)&=\frac{z e^{i[x-y-z]}}{(z-x) (y+z)}-\frac{z e^{i[-x-y+z]}}{(x+y) (z-x)}\,,\\
\hat g_2(p_0,p_1,p_2)&=\frac{z e^{i[-x+y-z]}}{(x-y) (x+z)}\,.
\end{align}
In verifying that the cocycle is regular it is important to use the Plucker identities. At the end of the day the two methods outlined above are similar to each other and the problem is to a find a function of three variables with specific properties.

\section{Higher-Spins on Background of Their Own}
\label{app:fluctuations}
As it was discussed in Section \ref{sec:UnfldHochReal}, having the Hochschild two-cocycle one can immediately write down the equations that describe propagation of higher-spin fields over a background of their own that is represented by a flat connection of the higher-spin algebra. For the case of the $4d$ HS theory the relevant vertex can be found in \cite{Vasiliev:1988sa}. Changing slightly notation we have
\besubeqs\label{fluctapp}
\begin{align}
d\Omega&=\Omega\star \Omega\,, \\
d\omega&= \Omega\star \omega+\omega\star \Omega +V(\Omega,\Omega,C)\,,\\
dC&=\Omega\star C-C\star \tilde{\Omega}\,.
\end{align}
\esubeqs
Using the calculus of symbols of operators of Appendix \ref{app:symbols} one, for example, finds
\begin{align}
\Omega\star C&= e^{i[p_{01}+p_{02}+p_{12}+\pb_{01}+\pb_{02}+\pb_{12}]}\Omega(y_1,\bry_1) C(y_2,\bry_2)\Big|_{y_1,y_2,\bry_1,\bry_2=0}\,,
\end{align}
where $p_{ij}\equiv p_i \cdot p_j$. The vertex $V$ consists of two elementary vertices, one for $y$ and another for $\bry$:
\begin{align}\label{fourdvert}
\hat V(p_0,p_1,p_2,p_3;\pb_0,\pb_1,\pb_2,\pb_3)&=e^{i\theta}\hat v(p_0,p_1,p_2,p_3;\pb_0,\pb_1,\pb_2,\pb_3)+e^{-i\theta}\hat v(\pb_0,\pb_1,\pb_2,\pb_3;p_0,p_1,p_2,p_3)
\end{align}
that act on $\Omega(y_1,\bry_1) \Omega(y_2,\bry_2)C(y_3,\bry_3)$. The elementary vertex is
\begin{align*}
&\hat v(p_0,p_1,p_2,p_3;\pb_0,\pb_1,\pb_2,\pb_3)=\\
&\qquad\qquad \qquad=p_{12} e^{i\left[\pb_{01}+\pb_{02}+\pb_{03}+\pb_{12}+\pb_{13}+\pb_{23}\right]}\int e^{i\left[\left(1-2u_1\right) (p_{01}-p_{13})+\left(1-2u_2\right) (p_{02}-p_{23})+\left(1+2u_1-2u_2\right) p_{12}-p_{03}\right]}\,,
\end{align*}
where the integration is over the $2d$ simplex ${0\leq u_1\leq u_2\leq1}$ and the vertex represents $\Phi(\Omega,\Omega)\star \tilde{C}$. Using the pure gauge representation for $\Omega$, $\Omega=-g^{-1}\star dg$, where $g=g(y,
\bry|x)$ can be multivalued, we reduce the equations to
\begin{align}\label{twoform}
d\omega_0&=g\star \mathcal{V}(g^{-1}\star dg,g^{-1}\star dg,g^{-1}\star C_0\star \tilde{g})\star g^{-1}\,, & dC_0&=0\,.
\end{align}
Let us note that in the case of higher-spin fields in $3d$ the equations are much simpler:
\begin{align}
dC&= A\star C-C\star B\,, && dA=A\star A\,, && dB=B\star B \,,
\end{align}
where $A$ and $B$ are two one-forms and $C$ is a zero-form, all taking values in the higher-spin algebra $\mathfrak{hs}(\lambda)$, \cite{Vasiliev:1989re, Feigin}. The equations describe a scalar field on a higher-spin background defined by $A$ and $B$, see also \cite{Vasiliev:1992gr}. The simplicity is due to the fact that there are no Weyl tensors for massless fields with spin $s=2,3,...$.

Higher-spin fluctuations have to obey \eqref{fluctapp} in any $d>3$. In $d>4$ the Hochschild cocycle is known implicitly from \cite{Vasiliev:2003ev}: it is the product of the $A_1$ cocycle and a pure star-product with respect to some additional oscillators $y^a_\ga$ with certain factorization over an ideal required to get its action on the higher-spin algebra.

\section{$Sp(4)$ Cocycle}
\label{app:spfour}
As an illustration and in view the possible importance of the $A_2$ Hochschild cocycle for the $4d$ higher-spin theory let us evaluate the integrals over the simplex in \eqref{ffscocycle}. We use the FFS formula directly, which should correspond to some gauge in equations \eqref{justcocycle}, the advantage being that the FFS cocycle vanishes on $sp(2n)$. The answer is convenient to write in terms of
\begin{align*}
x&=p_{01}+p_{02}-p_{03}-p_{04}+p_{12}-p_{13}-p_{14}-p_{23}-p_{24}+p_{34}\,,\\
y&=p_{01}-p_{02}-p_{03}-p_{04}-p_{12}-p_{13}-p_{14}+p_{23}+p_{24}+p_{34}\,,\\
z&=-p_{01}-p_{02}-p_{03}-p_{04}+p_{12}+p_{13}+p_{14}+p_{23}+p_{24}+p_{34}\,,\\
u&=p_{01}+p_{02}+p_{03}+p_{04}+p_{12}+p_{13}+p_{14}+p_{23}+p_{24}+p_{34}\,,\\
v&=p_{01}+p_{02}+p_{03}-p_{04}+p_{12}+p_{13}-p_{14}+p_{23}-p_{24}-p_{34}\,,
\end{align*}
\begin{align}
\hat\Phi=\det|p_1,p_2,p_3,p_4|\times \Biggr[&\frac{e^{ix}}{(u-x) (v-x) (x-y) (x-z)}+\frac{e^{iy}}{(u-y) (y-v) (y-x) (z-y)}+\notag\\
+&\frac{e^{iz}}{(u-z) (v-z) (x-z) (y-z)}+\frac{e^{iu}}{(u-v) (u-x) (u-y) (u-z)}+\\
+&\frac{e^{iv}}{(u-v) (v-x) (v-y) (z-v)}\Biggr]\notag\,.
\end{align}
Thinking of the AdS/CFT applications we should evaluate it on the boundary-to-bulk propagators for $\omega$. First of all we see that $\Phi$ is local, i.e. $\Phi$ contains a finite number of derivatives, provided the spins of the four arguments are fixed. The next correction to it, which is of order $\omega^4C$, is also local. The $\omega$-propagators\footnote{The propagators were found in \cite{Giombi:2009wh} in a more complicated form, but it can be shown that they can be cast into the simple form discussed here. E.S. is grateful to S.Didenko with whom this result was obtained.} are of the schematic form $h \exp[i y\cdot\chi]$, where $h$ is the vielbein, whose indices are contracted with polarization spinors introduced in \cite{Didenko:2012tv}. The most important fact is that they are simple exponential functions. Therefore, the vielbeins eventually make a volume form, while the arguments $p_i$ can be directly identified with $\chi$. The cocycle starts to be effective from $s=2$ due to the determinant factor, but due to the fact that the Hochschild cocycle can be made to vanish when one of the arguments is in $sp(4)$, i.e. it is $sp(4)$-basic, it starts with $s=3$. Therefore, all of the arguments should correspond to fluctuating fields: if one of them is replaced by the $AdS$-background the cocycle vanishes identically. It is worth emphasizing again that the expansion scheme here is in powers of zero-forms $C$ and counting the orders in the weak-field expansion scheme is different: the full quartic on-shell action\footnote{See \cite{Bekaert:2014cea,Bekaert:2015tva} for a part of the quartic on-shell action within the weak-field expansion scheme.} in the weak-field expansion requires the knowledge of the terms of order $\omega C^k$, $k=0,1,...,4$. The same time, each of these terms contributes to infinitely many orders in the weak-field expansion. For example, one good candidate is the quantum trace \cite{FFS}:
\begin{align}
    \mathrm{Tr}(f)=\int \mathrm{str}[\Phi(\omega,\omega,\omega,\omega)\star f]\,,
\end{align}
where $f$ is covariantly constant with respect to $\omega$, e.g. Killing tensor.

\end{appendix}

\setstretch{1.0}
\bibliographystyle{utphys}
\bibliography{megabib.bib}

\end{document}